\documentclass[aip,graphicx]{revtex4-1}

\draft 

\RequirePackage[normalem]{ulem}  
\usepackage{hyperref}
\usepackage{nameref}

\usepackage{graphicx}
\usepackage{dcolumn}
\usepackage{bm}

\usepackage{xcolor}
\usepackage[utf8]{inputenc}
\usepackage[T1]{fontenc}
\usepackage{mathptmx}
\usepackage{amsmath}

\begin{document}

\title{A Framework for Causal Discovery in non-intervenable systems} 



\author{Peter Jan van Leeuwen}
\thanks{Corresponding author}
\email[]{peter.vanleeuwen@colostate.edu}
\altaffiliation{Also at University of Reading, Reading, RG6 6BB, UK}
\affiliation{Colorado State University, Fort Collins, 80523-1371, USA}

\author{Michael DeCaria}
\email[]{Michael.DeCaria@colostate.edu}
\thanks{}
\affiliation{Colorado State University, Fort Collins, USA}
\author{Nachiketa Chakraborty}
\email[]{n.chakraborty@reading.ac.uk}
\thanks{}
\affiliation{University of Reading, Reading, Reading, RG6 6BB, UK}
\author{Manuel Pulido}
\email[]{pulido@exa.unne.edu.ar}
\thanks{}
\affiliation{Department of Physics, Universidad Nacional del Nordeste, Corrientes, 3400, Argentina}

\date{\today}

\begin{abstract}
Many frameworks exist to infer cause and effect relations in complex nonlinear systems but a complete theory is lacking. A new framework is presented that is fully nonlinear, provides a complete information theoretic disentanglement of causal processes, allows for nonlinear interactions between causes, identifies the causal strength of missing or unknown processes, and can analyze systems that cannot be represented on Directed Acyclic Graphs. The basic building blocks are information theoretic measures such as (conditional) mutual information and a new concept called certainty that monotonically increases with the information available about the target process. The framework is presented in detail and compared with other existing frameworks, and the treatment of confounders is discussed. While there are systems with structures that the framework cannot disentangle, it is argued that any causal framework that is based on integrated quantities will miss out potentially important information of the underlying probability density functions. The framework is tested on several highly simplified stochastic processes to demonstrate how blocking and gateways are handled, and on the chaotic Lorentz 1963 system. We show that the framework provides information on the local dynamics, but also reveals information on the larger scale structure of the underlying attractor. Furthermore, by applying it to real observations related to the El-Nino-Southern-Oscillation system we demonstrate its power and advantage over other methodologies.
\end{abstract}


\maketitle 
\section*{Lead Paragraph}
\textbf{Unraveling cause and effect in complex systems is one of the fundamental tasks of science. This becomes a considerable challenge in systems where interventions are not possible and our only sources of information are time series of the processes of interest. Huge progress has been made for systems in which the underlying causal structure can be represented on a standard graph, in which each process is represented by a node, and causal links by arrows from one node to another. However, there are many systems where the causal structure is too rich to be represented on such graphs. We developed the first complete causal discovery network for such systems, decomposing the causal influence of each driver into its direct contribution to a target process and its contribution with any other driver, any two other drivers, etc. Furthermore, we are able to quantify the influence of unknown driver processes, so that we know how accurate our causal decomposition is. The usefulness is demonstrated via many examples. The new framework allows for new insight in complex systems for which often only time series are available, such as the atmosphere and the ocean and climate, astrophysics, the human brain, etc. }

\section{Introduction}
Causal discovery can be roughly divided in four different tasks: uncovering the causal network, building structural causal models, studying the influence of interventions, and couterfactual reasoning. This paper deals with the first task: uncovering the causal network with a goal to progress scientific knowledge of a system. The information source are time series of variables or processes of the system of interest. 

Since the systems of interest are highly nonlinear, building structural equation models is a difficult task that we will not consider here. We refer to the literature on e.g. Relevant Vector machines, Bayesian symbolic regression and many other developments for interesting progress in that field, including using neural networks as structural equations, see e.g. \cite{Bishop2006,Jin2020, Shojaie2015, Glymour2019, Matthew2021}.
We do not consider interventions because the systems of interest do not allow for interventions, either because interventions are impossible, unethical, or would change the dynamics of the system such that the intervention studies a completely different structure. 
Examples of application are systems where the internal dynamics is so complex that manipulating external forcings does not reveal much about the internal dynamics. One can think about systems such as the atmosphere or the ocean, biological food webs, complex chemical systems, the brain, etc. 
Interventions on internal variables of these systems typically push these systems off their attractor, resulting in causal inference on regions of state space that are not of interest. 
One could argue that a small enough intervention, small in magnitude, spatial and temporal extent, would still be useful in such systems. However, due to the strong feedbacks it is hard to infer what the direct influence of the intervention is, and what is related to reactions to the feedbacks. For this reason we will call such systems non-intervenable.

In fact, excluding interventions means we cannot use Pearl and coworkers' beautiful do-calculus, in which the 'do' operator means a direct intervention. If a given causal network allows interventions, one can then use do-calculus to calculate the effect of the intervention without actually doing the intervention\cite{Pearl2009}.
While our framework potentially complements do-calculus, showing such is beyond the scope of this paper.
Our intended applications do not allow either actual or theoretical interventions, so we will use the term 'non-intervenable', or 'observational causal inference', to denote our path of study,
restricting ourselves to use data only from observations, reliable simulations, or a combination. (See \cite{Bareinboim2016}, for example, who explore a form of importance sampling to explore one data set to infer causal inference on another.)

Because of these limitations we define causal relations between a target process $x$ and potential driver processes $y$ through two criteria: 1) the cause $y$ precedes the effect $x$, and 2) a causal relation between processes or variables in a system exists if there is flow of information between them, hence information flow from $y$ to $x$. Our goal is not to predict the future of $x$ from $y$; that would be a next step. The goal is to increase understanding of a system by establishing how information flows through the system, where information is to be interpreted as reduction of uncertainty.

Precise mathematical descriptions of observational causal inference started with the seminal works of Wiener \cite{Wiener1956} and Granger \cite{Granger1969}
in the 1950's and '60's. Their basic idea was to build a minimal structural equation model by defining a set of functions from observed variables and determine the regression coefficients of these driver functions, or driver processes, on a target process. 
A large regression coefficient suggests a large causal influence of that driver process on the target process. 
If the regression coefficient of a process is small that process is not considered a cause for the target process. Pruning in this way leads to a minimal model 
and this minimal model is then the causal model of the target process. 
In this framework, one has to define the potential driver processes directly, or nonlinear functions of them, beforehand, and the causal inference is in essence looking for linear cross-correlations between linear or nonlinear functions.

Granger causality is based on the idea that a process is a driver of a target if it reduces the unexplained variance in the target process. The driver can be nonlinearly related to the target, but in that case the functional form of the relation has to be specified. Methods based on information theory avoid the specification of this functional form by considering the reduction in unexplained entropy in the target process. The first example of this kind is transfer entropy \cite{Schreiber2000}, and many extensions
are now available. The idea here is to define the causal strength of a process $y$ to a process $x$ as the conditional mutual information $I(x_t;y_{<t}|z_{<t})$ in which $z$ represents all other processes, and $y$ and $z$ are in the past of $x_t$. These methods pursue the identification of the causal network, but are not useful to build an actual structural equation model, because information theoretic measures such as (conditional) mutual
information are invariant under single-variable nonlinear monotonic transformations.
Hence, these methods cannot distinguish between a model in which a variable $x$ is present, or say $\exp(x)$. 

Several of these methods rely on graphical representations, and algorithms typically start 
either from an empty graph and add strong relations, or from a fully connected graph and prune weak relations, until a minimal unidirectional acyclic graphical model is found that represents the causal network. An example of building up from an empty graph is the Peter and Clark (PC) algorithm \cite{Spirtes1991}, and the so-called Greedy equivalence Search \cite{Chickering2002} is an example of a pruning algorithm. The strength of relations is determined via conditional independence tests (e.g. PC)  or via scoring rules (Greedy equivalence Search), and the emphasis is more on establishment or removal of causal links, and hence determining the causal structure, rather than determining the actual causal strength (defined in whatever way). Recently, Sun et al.\cite{Sun2014} pose the problem as an information theoretic optimization method.
 
These methodologies have been extended to high dimensions and in particular applied to earth system processes by e.g.  \cite{Runge2015a, Runge2015b,Runge2015c}. Many other formalisms have been proposed, and the excellent reviews of Runge et al. \cite{Runge2019} and Glymour et al. \cite{Glymour2019} contain much of present-day efforts for systems in which interventions are not possible. As mentioned, our interest is in those kind of systems.

More general methods to generate causal models for non-interveneable systems have been developed since. For instance, Convergent Cross Mapping \cite{Sugihara2012} tries to find the underlying dynamical system using Takens' embeddings. The idea is that if a {\em driver} variable can be predicted from the time embedding of the {\em target} process, then that driver process is a cause of the target process. The reasoning is opposite to Granger causality, which tries to predict the {\em target} from the {\em drivers}. 
Sugihara et al \cite{Sugihara2012} provides a very careful discussion of the connection of Convergent Cross Mapping and Granger causality. Recently in Leng et al. \cite{Leng2020}, an extension called Partial Cross Mapping was proposed that can distinguish direct (i.e. $y \rightarrow x$) from indirect (i.e. $y \rightarrow z \rightarrow x$).   
Unfortunately, these CCM-based methods are less suited when the underlying process is strongly stochastic, or heavily corrupted by unknown processes ('noise'), because the embedding methodology is not robust to the presence of noise. Furthermore, each causal inference based on CCM looks for the behaviour of binary connections, and it cannot infer how different causes work together to influence the target, which, as explained below, is a crucial motivation for the development of our methodology.

A recent surge of causal discovery methods originated in machine learning. The typical assumption is that information is available on interventions in a complex system, and large data sets are used to infer the average treatment effect, for instance by including causal regularizers, see e.g. \cite{Guo2020} and references therein for the many studies in these areas. Many studies search for causal features out of a finite set of features, while our focus is on processes where the causal feature set is continuous, and hence infinitely large.
The standard assumption is that the underlying processes can be represented on a standard Directed Acyclic Graph (DAG) as this is the underlying structure of a typical neural network, while, as argued below, graphs are too restrictive for the causal structures we are interested in. Machine-learning has also been used to infer the influence of confounders on the causal net. The idea is that, although the confounder itself is not known, we do have proxy variables from which representations of the confounder influence can be estimated, for instance via Causal Effect Variational AutoEncoders \cite{Louizos2017, Tran2015}. 

One issue with the methods discussed above is that there are many examples in the real world where causes are nonlinear interactions between driver processes, and the above transfer-entropy-based and other methods cannot disentangle this properly, see e.g.  \cite{James2017,Runge2015a}. A simple example  is a transistor in which one process acts as a gate keeper for the connection between other processes. However, there is a more fundamental limitation of these methodologies.

All methods mentioned above can be represented on standard graphical networks, such as Bayesian networks or Markov-random fields, called causal graphs. Most theory and methodologies are based on so-called Directed Acyclic Graphs (DAG's), in which each edge is an arrow, and there is at most one arrow between two vertices. These graphs represent the underlying joint probability of the system. However, these networks are designed to represent dyadic, so binary, interactions between the variables while in many systems the interactions are polyadic \cite{James2016, Runge2018}. A simple example demonstrates the issue. Assume that driver $y=0$ if $z=0$ and $y$ can take on 0 or 1 with equal probability when z = 1. The target $x=yz +1 -z -y + \eta_x$, in which $\eta_x$ denotes random noise that is independent of $y$ and $z$. All conditional mutual information terms of $x$ with any driver conditioned on the other driver is zero, so a graph representation will consist of three nodes without edges. However, there is a nontrivial relation between the variables. One could argue that the problem is rather special in that it doesn't satisfy the Faithfulness and Causal Markov \cite{Spirtes2000, Pearl2009} conditions that are typically assumed in causal discovery, which together state that independency on the DAG means independency in the joint pdf, and vice versa. (See also the discussion in \cite{Weinberger2018}) However, the problem runs deeper. 
The issue is that many joint pdfs cannot be represented on a DAG and hypergraphs are needed. 

\begin{figure}
\centering
\includegraphics[width=6 cm]{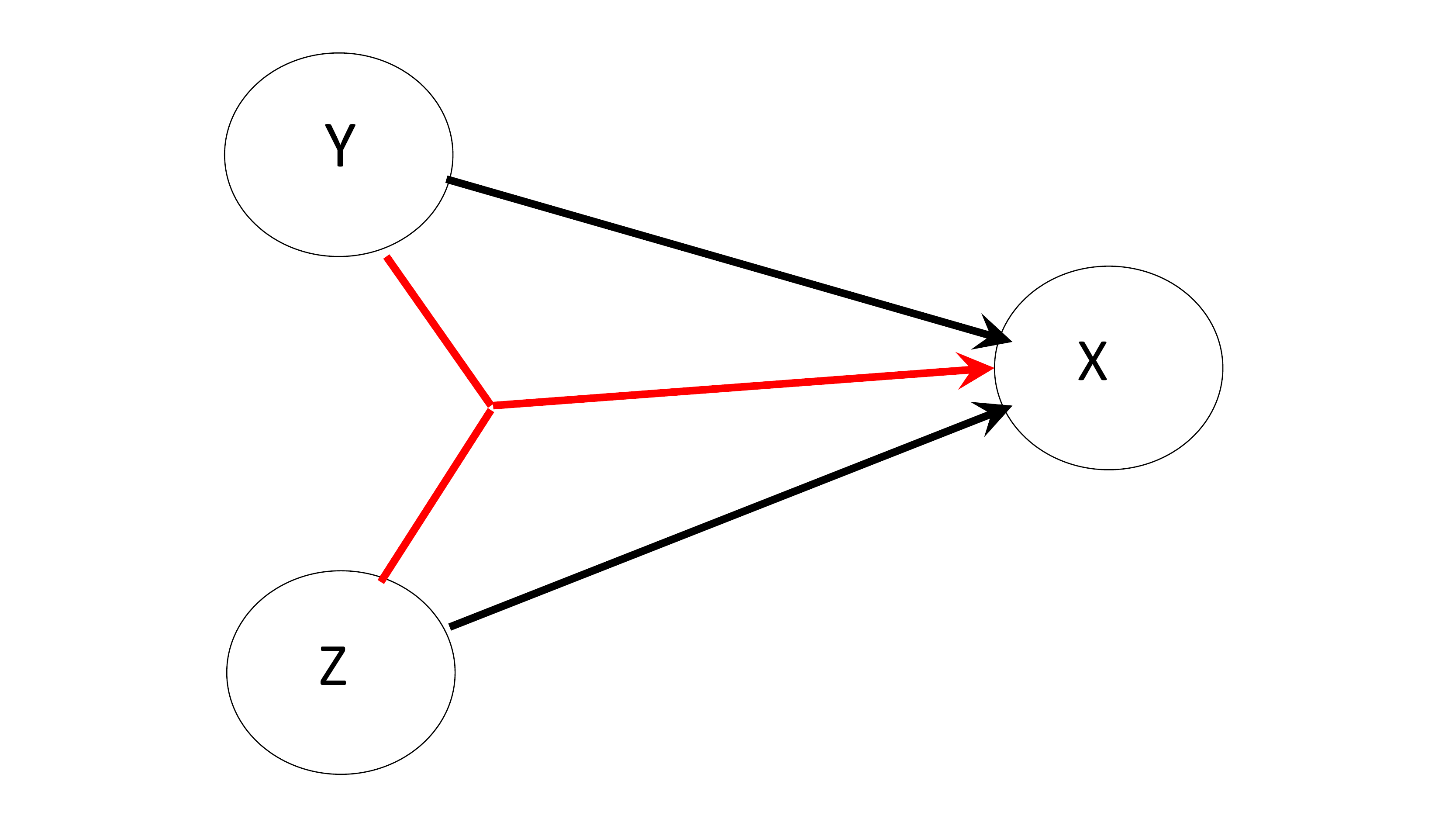}
\caption{Attempt to put the structure $x=yz+y+\eta_x$ on a graphical model. The black arrows denote the binary connections between $y$ and $x$, and $z$ and $x$. The red arrow shows the combined influence of $y$ and $z$ on $x$, the $yz$ term. This link is present but cannot be represented on the graph which allows only one edge between nodes, and edges are not allowed to combine or split. }
\end{figure}   
Let us try to put this structure $x=yz + y + \eta_x$ on a graphical model, as in figure 1. The problem is that the combined influence of $y$ and $z$ cannot represented by one edge, but edges need to be allowed to merge, for which a hypergraph is needed. Many processes in nature are of the nonlinear form depicted in figure 1, for instance advection terms in fluid dynamics, the interaction of radiation with matter, predator-prey biological systems, chemical reactions etc. Standard graphical networks such as DAGs are not general enough to represent these processes. Our framework is not based on such graphical models, and in fact is developed with this kind of polyadic interactions in mind.

A framework that can handle these polyadic interactions, at least in principle, is the interesting contribution by Williams and Beer \cite{Williams2010}. They introduced a nonnegative decomposition of multivariate information, the so-called Partial Information Decomposition (PID)
that does allow for the inclusion of joint information, so it does allow for polyadic interactions.
The basic idea is that the total driver process information can be split into unique contributions $U$ from each driver, 
synergistic contributions $S$ with other driver processes, and redundant contributions $R$. Redundant contributions are contributions to the target process that 
two or more driver processes have in common. These descriptions are rather vague, which allows for freedom, but also hampers applicability. For a system with only three processes, one target $x$ and
two driver processes $y$ and $z$ the mutual information between the target and the drivers is decomposed as
\begin{eqnarray}
I(x;y,z) & = & U(x;y|z) + U(x;z|y) + S(x;y,z) +R(x;y,z) \nonumber \\
I(x;y) & = & U(x;y|z) + R(x;y,z) \nonumber \\
I(x;z) & = & U(x;z|y) + R(x;y,z)
\label{eq:PID}
\end{eqnarray}
This system consists of 3 equations for the 4 unknown contributions and hence is underdetermined. 
The only general condition is that all 4 quantities have to be non-negative. 
We can eliminate the unique contribution $U$ by forming
\begin{equation} 
I(x;y,z) - I(x;y)-I(x;z) = S(x;y,z)-R(x;y,z)
\end{equation}
and hence the difference between $S$ and $R$ is defined in terms of mutual informations, but not each term individually.
In information theory this combination of mutual informations is minus the interaction information, and hence:
\begin{equation} 
I(x;y;z)  = R(x;y,z)-S(x;y,z)
\label{eq:RminS}
\end{equation}
which can have any sign. Furthermore, from the basic PID equations and the conditional information relation $I(x;y|z)=I(x;y,z)-I(x;z)$ we can derive 
\begin{equation} 
I(x;y|z) = U(x;y|z) + S(x;y,z)
\label{eq:UplusS}
\end{equation}
showing that the conditional mutual information is interpreted as the sum of the unique and synergistic information in the PID framework.

Many definitions have been explored defining one of the variables in the PID framework and deducing the others from the framework, but all 
have their weaknesses. 
For instance, Barrett et al \cite{Barrett2015} showed that for dependent Gaussian source processes three popular interpretations of PID
 \cite{Williams2010}, \cite{Griffith2014} and \cite{Bertschinger2014}, and \cite{Harder2013} all lead to the situation that the weakest source 
 has zero unique contribution. However, this result is problematic.  Suppose that we do have a weakest source process that has unique information on the target process,
 in the sense that it contains information on the target that none of the other driver processes have. 
All three PID interpretations mentioned above insist that that unique contribution is zero, leading to a logical inconsistency.
 
Besides this, it is not clear if a unique contribution between a driver and target can be well defined in the first place. It can when the system can be decomposed on a graphical network, and in that case conditioning out other processes as in transfer entropy logically provides the unique contribution. However, when a cause consists of nonlinear interactions between drivers, which is reality for many systems in the natural world, conditioning is insufficient to define a unique contribution. As mentioned above conditioning can open gateways that are otherwise closed. The unique contribution of process $y$ on $x$, $U(x;y|z)$ is supposed to mean something like 'when driver $z$ is not present'. However, $z$ is always present. We would need to find measures that exclude all influence of $z$, but a general way to do that does not exist (unless we allow for interventions). These problems, to us, seem to point to serious issues with present-day interpretations of the PID formalism. 
That does not mean that the formalism is not useful, just that more work on its interpretation is needed.

In this paper we provide a new causal discovery framework for non-intervenable systems that is unique in several ways.
It does not rely on causal graphs, and hence its application is not restricted by the issues discussed above.
It is based on the notion of {\em certainty} instead of entropy.
Certainty increases monotonically with the amount of information we have about a target process. Another reason to introduce this new concept is that it is always nonnegative for unbounded variables, while differential entropy can be negative, obscuring clear interpretations. 
The total mutual information of all driver processes with the target process is interpreted as the increase of certainty compared to having only the time series of the target process.
We decompose the total mutual information in direct contributions from each driver to the target, and joint contributions between 2
processes, between 3 processes etc. 
We normalize the contribution of each process and define direct, joint, and total causal strengths from one process to another.
By normalizing each contribution different studies can be compared, and the certainty from the original time series of the target process, the so-called self certainty, 
can be reinterpreted as the contribution from unknown processes.
Hence we can quantify the contribution of unknown processes (such as 'noise'), and show that including new processes can only
decrease this contribution from unknown processes. This is different from confounder influence estimation in machine learning, in which proxies for the confounder influence have to be present, while we need no extra information on the confounders.

The paper is organized as follows. In the next section the basic ingredients of the new framework are introduced, followed by an
example of how to decompose the mutual information when 3 processes are involved. Then we show in section 4 
the general theory of the decomposition of the total mutual information, discuss confounders in section 5 and apply the framework to several examples in section 6.
The paper is finalized by a discussion and concluding section.

\section{Basics of the framework}
The problem we want to solve is to identify the relative influence of a set of nonlinearly interacting random processes $y_i$, $i=1,2,...,N$ on a target random process $x$. 
The subindex $i$ denotes a separate process, not a time index. Each of these processes themselves is a time series.
Relative influence is defined as 
the extent to which process $y_i$ increases our knowledge about process $x$. Hence we want to decompose
our predictive knowledge about $x$ in its contributions from all processes $y_i$ (which can include the past of any driver or even the past of $x$ itself),
written symbolically as:
\begin{equation}
(y_{1:N} \rightarrow x) = \sum_{i=1}^N (y_i \rightarrow x)
\end{equation}
where $(y_i \rightarrow x)$ contains all direct and joint contributions to $x$ in which $y_i$ is involved. We will show in sections 3 and 4 that this decomposition can indeed meaningfully be made with a proper definition of the arrow, even when the interaction of drivers with each other to influence the target is inseparable.

The full framework consists of a set of tools and operations as defined below. The framework defines a {\em causal web} that consists of driver nodes and a target node, and directed edges (arrows) pointing into the target. An arrow can directly connect a driver node to the target or it can merge with one or more arrows from other driver nodes. 
Figure 1 is the simplest example, but more complex examples are provided later. 

Each arrow that points into the target has a {\em causal strength} attached to it that describes the relative strength of the causal link between the one node it originates from, or the relative causal strength due to the interaction of several nodes that the arrow originates from. The causal strength are {\em normalized} to enhance understanding and comparison with other systems. The following subsections describe how these relative causal strength are calculated and how they should be interpreted. One special directed edge points into the target from unresolved processes. This can be noise, or other physical processes that could be of interest but we forgot to include in the driver set, the so-called confounders. The relative causal strength attached to this arrow allows for {\em confounder detection}, as explained in the chapter on confounders. 

The causal strength calculations are based on {\em mutual information and conditional mutual informations among drivers conditioned on other drivers}, as detailed below. In principle the strength of every possible interaction between drivers towards the target needs to be calculated to obtain a full description of the causal structure of the system. Our method is minimally causally sufficient in accordance with definitions in \cite{VanderWeele2009}. It is sufficient because for every target variable, the set of mlinks represents all possible causal influences due to combination of two or more drivers. And it is minimal as any subset of these would not be causally sufficient in general.

\subsection{Entropy and mutual information}
The time-lagged mutual information $I(x;y_{1:N})$ between a target process $x$ and a possible driver process $y$, or a whole range of driver processes $y_{1:N}$ is defined via the Shannon entropy $H(..)$ as 
\begin{equation}
I(x;y_{1:N}) = H(x)  - H(x|y_{1:N}) 
\label{eq:basic}
\end{equation}
where we assume a positive time lag between process $x$ and driver processes $y_{1:N}$ for a causal link.

This lagged mutual information denotes the reduction in entropy of process $x$ when we condition on the processes $y_{1:N}$.
We want to interpret the entropy in terms of information as in Shannon's entropy, but we are interested in
the case that each process lives on an unbounded domain.  The differential entropy defined as 
\begin{equation}
H_{diff}(p) = - \int p(x) \log p(x) \;dx,
\end{equation}
where $p(x)$ is the probability density function (pdf) of a process $x$, cannot be used because it can be negative and it is not invariant under nonlinear transformations of the variables. 
Instead, we use the relative entropy, relative to a reference process with probability density $q(x)$ as:
\begin{equation}
H_{rel}(p||q)
= \int p(x) \log \left[ \frac{p(x)}{q(x)} \right]  \;dx
\label{eq:relentropy}
\end{equation}
The relative entropy is positive for any choice of $q(x)$, as long as its support is equal to or larger than that of $p(x)$.
This density $q(x)$ will provide an offset relative to $p(x)$, the pdf of the process of interest.
Although this offset density cancels in equation (\ref{eq:basic}), it influences the size of our causal strengths between processes. 

Many choices can be made for this reference density. Ideally, it has as little structure as possible, such as a uniform density. However, a uniform density does not exist on an unbounded domain. In this paper we will mainly use the Cauchy or Lorentz pdf, given by:
\begin{equation}
q(x)=\frac{1}{\pi } \frac{\gamma}{\gamma^2+(x-\mu_x)^2}
\end{equation}
defined by a  with width parameter $\gamma$ and mean $\mu_x$. A logical choice for the mean of the pdf is the sample mean. We choose $\gamma$ such that the reference density has the same entropy as the density with maximum entropy based on the mean and variance of the original process, so as the Gaussian. Since the entropy of the Cauchy distribution is $\log (4 \pi \gamma)$ and that of the corresponding Gaussian $(1/2) \log (2 \pi e \sigma_x^2)$ we can identify $\gamma = \sqrt{(e/8 \pi)} \sigma_x$, where $e$ is the base of the natural logarithm. With this choice the reference density can be interpreted as a maximum entropy pdf in the sense that it has the same maximum entropy as a Gaussian, but on top of that it has infinite variance.

Other choices can be used too, e.g. a Gaussian with sample mean and variance, or a uniform pdf with boundaries defined by the sample minimum and maximum values. We will discuss the influence of the reference density in section 6.3. 

\subsection{Certainty as information theoretic measure}

In the previous section we introduced the relative entropy as an important quantity in our framework. We now introduce a related quantity called {\em certainty}, defined as:
\begin{equation}
W(x |y_{1:N}) =  \int p(x,y_{1:N}) \log \left[\frac{p(x |y_{1:N})}{q(x)}\right] \;dx dy_{1:N}
\end{equation}

This quantity is a relative entropy with a reference density that still has to be determined.
We also introduce the unconditioned version, called the {\em self-certainty} in this context, as: 
\begin{equation}
W(x) =  \int p(x) \log \left[\frac{p(x)}{q(x)}\right] \;dx
\end{equation}
which is also a relative entropy. The reference density is an important quantity for the size of the noise term, and hence is related to confounder detection. It should be chosen as uninformative as possible, so uniform for discrete variables (which is not the focus of this paper), and a wide distribution, wider than the target, for continuous variables.

We have  $0 \leq W(x) \leq \infty$, 
with boundaries attained when $p(x )= q(x)$ or a delta Dirac function, respectively. 
This is in contrast to entropy, which is a measure of 
uncertainty. Hence, $W$ can be seen as a measure of certainty: given a wide reference density, the narrower the pdf of $x$, the more we know about $x$, and indeed the higher our certainty
about $x$. Similarly, for the conditional variant we have $0 \leq W(x ) \leq W(x |y_{1:N}) \leq \infty$, as can easily be verified.

The certainty and the self certainty are related through the information theoretic relation
\begin{equation}
W(x |y_{1:N}) = W(x ) + I(x;y_{1:N})
\label{eq:basic1}
\end{equation}
as follows directly from the definition of the terms. 
Expression (\ref{eq:basic1}), which only contains non-negative terms, will be the basis for our causal inference.  
The term $W(x )$ denotes the amount of self-certainty we have on process $x$.
The mutual information term $I(x;y_{1:N})$ is 
the increase in certainty on $x$, due to knowledge of $y_{1:N}$. 
$W(x |y_{1:N})$ denotes the information we have on process $x$ when we condition on processes $y_{1:N}$, 
so when we know what these processes $y_{1:N}$ are doing.

The next section will introduce normalization, which will allow for a more direct interpretation of the terms in the
theory, and will make different experiments comparable.

\subsection{The need for normalization}
We can calculate the quantities above but they would have little direct meaning.
What does a mutual information of, say, 2.6 mean? Some meaning can be extracted if we compare what this value 
would mean for a standard process, such as a Gaussian, but if the process is far from Gaussian, e.g. multimodal,
this explains very little. 
Since our quantity of interest is the relative contribution to the certainty in $x$ brought by each process, we 
normalize (\ref{eq:basic1}) by the certainty conditioned on all these processes, $W(x|y_{1:N})$:
\begin{equation}
1 = \frac{W(x )}{ W(x |y_{1:N})} + \frac{I(x;y_{1:N})}{W(x |y_{1:N}) }
\label{eq:basic2}
\end{equation}
Using normalization by $W(x|y_{1:N})$ we find as the relative influence of all processes $y_{1:N}$ 
on process $x$, or the {\em causal strength of processes $y_{1:N}$ towards process $x$}:
\begin{equation}
cs(x;y_{1:N}) = \frac{(y_{1:N} \rightarrow x)} {W(x |y_{1:N})} = \frac{I(x;y_{1:N})}{W(x |y_{1:N})}
\end{equation}
and hence
\begin{eqnarray}
1 = cs(x)& = &  cs(x;y_{1:N}) + cs(x;x) \nonumber \\
& = & \frac{I(x;y_{1:N})}{W(x |y_{1:N})} + \frac{W(x)} {W(x |y_{1:N})} 
\label{eq:normalized}
\end{eqnarray}
This last equation is the same as (\ref{eq:basic2})
showing the contributions to $x$ by processes $y_{1:N}$ and its self-certainty. The importance of the normalization is that now we can compare different studies on causal discovery. Instead of having to infer if a mutual information of say 2.6 is large or not, we know immediately if a causal strength of say 1/2 is large as this means that that process contributes $50\%$ to explaining the target process.

There is, however, another reason for introducing normalization.
To understand the framework further we assume that the underlying equation that governs process $x$ can be written as
\begin{equation}
g(x,y_{1:N},\eta)=0
\end{equation}
for some function $g(..)$, in which $\eta$ denotes all processes not included in $y_{1:N}$, so all unresolved or unknown processes
that are typically considered as noise. This assumption is completely general. The process $\eta$ is included because any real world time series will always contain unknown or unresolved processes as well as observation noise, so process $\eta$ does play a role in reality.
If we would know the process $\eta$ we could calculate $I(x;y_{1:n},\eta)$
and the result would be $\infty $. In that case $W(x)$ would be insignificant compared to the mutual information.
This suggests that the ratio between the self-certainty and the mutual information of the known processes $y_{1:N}$
gives us a measure of how close we are in taking all relevant processes for $x$ into account. 
This ratio contains the same information as the ratio between
$W(x)$ and $I(x;y_{1:n}) + W(x) = W(x |y_{1:N})$. 
This, then, suggests that the smaller $W(x)/W(x |y_{1:N})$ the more complete the processes $y_{1:N}$ are
in the causal description of $x$. 

To clarify this further, assume we discover a new important process $w$. 
Because $W(x |y_{1:N},w) = W(x|y_{1:N}) + I(x;w|y_{1:N})$ and $I(x;w|y_{1:N}) \geq 0$ because
it is a bivariate mutual information, we have $W(x |y_{1:N},w) \geq W(x|y_{1:N})$.
Since $W(x)$ does not change by incorporating $w$,
the ratio $W(x)/ W(x |y_{1:N},w)$ will be smaller than $W(x)/ W(x |y_{1:N})$. This means that the more relevant driver processes we include, the smaller the ratio between the self certainty and the certainty. {\em Hence, we can attribute this ratio to unmodeled processes}.
We thus find that the normalization by $W(x |y_{1:N})$ changes the interpretation of the $W(x)$ term from self information to the causal strength of unmodeled processes, and hence we identify $cs(x;x) = cs(x;\eta)$.
We consider this a very useful property that other frameworks lack.
\cite{McGill1954} develops this decomposition in the discrete setting and does notice that what we call self-certainty is related to what he calls noise. However, using this as a measure on the accuracy of the causal discovery is new. In section V we discuss the important case when the missed processes contain important information on the causal structure, the so-called confounders.

\section{Decomposing mutual information when 2 driver processes are involved}

Now that we have defined the general framework a method to quantify the individual contributions $ (y_i \rightarrow x)$ is developed.
As an example of how we determine individual contributions to the target process $x$ we first study
the case of target process $x$ and two driver or source processes $y$ and $z$.
For completeness we note that each of these processes $y$ or $z$ could be process $x$ itself, but lagged in time.
Equation \ref{eq:basic1} for three processes reads:
\begin{equation}
W(x |y,z) = I(x;y,z) + W(x),
\label{eq:overall}
\end{equation}
and our task now is to decompose $I(x;y,z)$ into the contributions from $y$ and $z$.

The influence of each process on $x$ can be divided in two contributions: a contribution when
we fix the other process, which we will call the {\em 1link contribution}, and a correction to that.
That correction by $y$ and $z$ together, so a 2link contribution is often only partially taken into account. For instance, the situation depicted in Figure 1, where the red arrow denotes a nonlinear interaction between the two drivers,
is often ignored in the literature that base the causal structure on DAG's (e.g. \cite{Pearl2009}, \cite{Runge2015a,Runge2018}) because that structure cannot be represented on such a graph.
While ignoring this contribution might be useful for some systems, we will show
in the examples that the present generalization is necessary for a full description of the causal network.

The conditional 1link contribution of process $y$ is found by conditioning on all other processes, so on process $z$
in this case.
This means that we study the influence of $y$ on $x$ when the influence of $z$ has already been taken into account because it is given.
This 1link contribution can be 
quantified by the conditional mutual information of $y$ to $x$ given process $z$:
\begin{equation}
 (y  \rightarrow x)_{1link} = I(x;y|z) 
 =  \int p(z) \int p(x,y | z) \log \frac{p(x,y|z)}{p(x|z)p(y|z)} \;dxdydz 
\end{equation}
That the 1link can be considered the direct contribution of process $y$ on $x$ can be seen from the conditioned version of equation (\ref{eq:overall}):
\begin{equation}
I(x;y|z) = W(\hat{x}|y,z) - W(\hat{x}|z)
\end{equation}
so the increase in certainty of $x$ when $y$ becomes available, given that we know the influence 
of process $z$. Similarly, for the 1link contribution from process $z$ we find:
\begin{equation}
 (z  \rightarrow x)_{1link} = I(x;z|y) 
\end{equation}
The correction term of both contributions has to be related to the combined influence of $y$ and $z$ on $x$.
Since the full contributions of $y$ and $z$ should add up to $I(x;y,z)$, as shown in the previous section,
the correction term has to be:
\begin{equation}
 (y\rightarrow x)_{2link}=I(x;y,z) - I(x;y|z) - I(x;z|y)
\end{equation}
This is half the the total contribution of both processes, minus their conditional 1link contributions. 
If this term is positive it can be interpreted as the contribution of the combination of $y$ and $z$ not
contained in the conditional 1link contributions from $y$ to $x$ and from $z$ to $x$, which can be termed the 'synergy'. A more direct phrasing would be that this term provides a measure on how $y$ and $z$ enhance each others influence on $x$. 
On the other hand, when it is negative it can be seen as the 'redundant' information in the 
conditional information. A more direct phrasing would be that the two drivers hinder each others influence on $x$.
Since this contribution is purely combined, i.e. it only acts when both $y$ and $z$ are active, the symmetry between $y$ and $z$ in this term demands that
it must be divided equally between the two processes. Hence the total contribution 
from $y$ to $x$ becomes:
\begin{equation}
 (y  \rightarrow x)_{total} =  (y  \rightarrow x)_{1link} +  \frac{1}{2}(y  \rightarrow x)_{2link}=  I(x;y|z) 
+ \frac{1}{2} \left[ I(x;y,z) - I(x;y|z) - I(x;z|y) \right]
\end{equation}
Using the standard relation $I(x;y,z) =   I(x;z|y) + I(x;y) $ we find
\begin{equation}
 (y  \rightarrow x)_{total} = I(x;y|z) 
+ \frac{1}{2} \left[ I(x;y) - I(x;y|z) \right]
\end{equation}
The quantity between the brackets is known as the interaction information, defined as:
\begin{equation}
I(x;y;z) =  I(x;y) - I(x;y|z) = I(x;z)- I(x;z|y)
\end{equation}
 Interaction information measures the influence of a variable $z$ on the amount of information shared between $x$ and $y$, but it can do this in a non-intuitive way. 
 For instance, when $y$ and $z$ are enhancing each others influence on $x$, conditioning on $z$, so fixing $z$ can diminish this enhancement, so  $I(x;y) > I(x;y|z)$ and the interaction information is positive. On the other hand, $z$ can open a pathway between $y$ and $x$ that is not present without $z$. In that case one would expect $I(x;y) < I(x;y|z)$, so the 2link is negative. The fact that the 2link can be negative shows that one cannot identify the 1link with 'unique' information from $y$ to $x$ in a PID framework interpretation, because non-unique information should still be positive.
For completeness, the total contribution from $z$ is:
\begin{equation}
 (z  \rightarrow x)_{total} = I(x;z|y) 
+ \frac{1}{2} \left[ I(x;z) - I(x;z|y) \right]
\end{equation}

Now we find the causal strength of $y$ to $x$ as:
\begin{eqnarray}
 cs(x ;y)  & = &    \frac{(y \rightarrow x)_{total}}{W(x|y,z)}  \nonumber \\
&  = & \frac{I(x;y|z)}{W(\hat{x}|y,z)}   +  \frac{1}{2} \frac{\left(I(x;y) - I(x;y|z) \right)}{W(\hat{x}|y,z)}
\label{eq:ytox}
\end{eqnarray}
and similarly for $z$.
The unmodelled or noise relative contribution to $x$ is given by:
\begin{equation}
cs(x;\eta) = \frac{W(\hat{x})}{W(\hat{x}|y,z)} 
\end{equation}
leading to the total causal strength towards $x$ as
\begin{eqnarray}
1 & = & cs(x ;y) +cs(x;z) + cs(x,\eta)   \nonumber \\
&  = & \frac{I(x;y|z)}{W(\hat{x}|y,z)}   + \frac{I(x;z|y)}{W(\hat{x}|y,z)}  + \frac{I(x;y;z)}{W(\hat{x}|y,z)} + \frac{W(\hat{x})}{W(\hat{x}|y,z)} 
\end{eqnarray}
As mentioned above, a large portion of previous literature on causal inference using standard graphs have systematically ignored the corrections to the 'pure' 1link contributions.
They thus missed potentially 
important parts of the causal network. It is true that the order of importance of processes $y$ and $z$ for $x$ will
not change when the 2link is included as that term is the same for $z$ and $y$. 
However, the ratio of the contributions will change. Furthermore, when more processes are present, 2links
(and higher order links) can change the order of importance compared to the 1link order, and
hence can lead to a completely different interpretation of the causal structure of the system.
We will see examples of this later.

We can make the link to the PID framework by using (\ref{eq:RminS}) and (\ref{eq:UplusS}), and decomposing our total contribution from $y$ to $x$ as:
\begin{eqnarray}
 (y  \rightarrow x)_{total} & = &  I(x;y|z) + \frac{1}{2} \left[ I(x;y) - I(x;y|z) \right] \nonumber \\
 &= &   U(x;y|z) + S(x;y,z) + \frac{1}{2} \left[ R(x;y,z)-S(x;y,z)\right]  \nonumber \\
 & = & U(x;y|z) + \frac{1}{2} S(x;y,z)+\frac{1}{2} R(x;y,z)
\end{eqnarray}
This suggests that the total contribution of $y$ to $x$ is a unique contribution and half the sum of the synergy and redundancy, all as defined in the PID framework. This makes sense if we invoke the symmetry argument that synergy and redundancy should only be included half for the $y$ contribution, with the other half for the $z$ contribution, but remember that none of the terms are defined uniquely in the PID framework. Given the difficulty in defining a unique contribution, our framework makes perhaps more sense than the PID framework. Our decomposition is based on the number and the identity of the 'active' (as opposed to conditioned on) variables in the mutual information, as explained further in the next section.

\section{Decomposing mutual information when N driver processes are involved}

When $N$ processes $y_i$, $i=1,2,...,N$ influence process $x$ we can generalize the above as follows.
To find the total contribution of each process $y_i$ we first quantify how much each of them contributes 
to $I(x;y_{1:N})$ on top of what all others contribute. 
Then we quantify how much each process contributes in combination only with one other process.
This is followed by how much each process contributes in combination only with two other processes, etc,
until we reach how much each process contributes in combination only with all other processes.
The word 'only' is important as we have to avoid double counting. 
This leads to a decomposition of the total contribution of process $y_i$ to $W(x|y_{1:N})$ as
\begin{equation}
 (y_i \rightarrow x)_{total}  = (y_i \rightarrow x)_{1link}  + \frac{1}{2} (y_i \rightarrow x)_{2links}
+ \frac{1}{3} (y_i \rightarrow x)_{3links}
+ ...  + \frac{1}{N}  (y_i \rightarrow x)_{Nlink}
\end{equation}
Factors such as  $1/2$ appear because each $2link$ process $y_i,y_j$ appears both in the contribution 
from $y_i$ and in the contribution from $y_j$. Hence, this contribution needs to be distributed between
these two process contributions. Since they both serve in equal capacity to this term each process contributes
1/2 of this term. A similar argument holds for all higher-link terms in this decomposition. The decomposition of the total mutual information in terms of all the links was also noted by \cite{McGill1954}, who developed the theory for discrete systems, and \cite{Brown2009}, who worked in arbitrary spaces but neither did decompose this further into the contributions from each driver separately, and hence neither did find our interpretation.

Each $mlink$ contains conditional mutual informations of the form $I(x;y_i,z| w)$, 
in which $z$ is a (m-1) subset of $y_{\neq i}$,
and $w$ contains those processes that are not process $y_i$ and not in $z$. 
This conditional mutual information contains all possible interactions between the active variables $y_i$ and 
all variables in $z$, including lower order links. To make sure this term only contains pure $mlinks$ we 
need to subtract all links of lower order, so $(m-1)links$, $(m-2)links$
etc all the way to the conditional $1links$, contained in the original $mlink$ set, to avoid double counting. An expression for the mlink of driver $y_i$ can be written as follows. First define $I_i$ as a set of $m-1$ non-overlapping indices from $\{1,...N\}$ that do not contain $i$, and $m_i$ the set $\{i,I_i\}$. Then define $m_{i-1}$ as a set of $m-1$ non-overlapping indices from the set $\{m_i\}$, and $m_{i-2}$ as a set of $m-2$ non-overlapping indices from the set $\{m_{i-1}\}$, etc. With these definitions the mlink of driver $i$ in an N-driver system can be written recursively as:
\begin{eqnarray}
(y_i \rightarrow  x)_{mlinks}  & = &  \sum_{all\;I_i} \left[ I(x;y_{j \in m_i}|y_{k \notin m_i}) \right. \nonumber \\
& &  -   \sum_{all \; m_{i-1}}  \left(y_{j \in m_{i-1}} \rightarrow x \right)_{(m-1)links}  ...  \nonumber \\
& & \left.  - \sum_{all \; m_{m-1}} \left(y_{j \in m_{m-1}} \rightarrow x \right)_{1links}  \right]
\end{eqnarray}

As an example, when 3 processes influence $x$ (N=3) we find, for each $i$:
\begin{eqnarray}
(y_i \rightarrow  x)_{3link}  & = &  I(x;y_1,y_2,y_3)  \nonumber \\
& &  -   \left (\hat{I}_{1,2|3}+\hat{I}_{1,3|2}+\hat{I}_{2,3|1}  \right) \nonumber \\
& & -   \left(\hat{I}_{1|2,3} + \hat{I}_{2|1,3} + \hat{I}_{3|1,2} \right)
\end{eqnarray}
in which the $2links$ are given by
\begin{equation}
\hat{I}_{i,j|k} = I(x;y_i,y_j|y_k) - \left(\hat{I}_{i|j,k} + \hat{I}_{j|i,k}  \right)
\label{eq:hat}
\end{equation}
and for the $1links$:
\begin{equation}
\hat{I}_{i|j,k} = I(x;y_i | y_j,y_k)
\end{equation}

Note the structure of this decomposition. Every term $I(x;y_i,y_j|y_k)$ contains two 1link contributions that need to be subtracted to define the 2link. This is similar to what we did in section 3, but now with the extra conditioning on $y_k$. For the 3link we have that every term $I(x;y_1,y_2,y_3) $ contains both 2links and 1links that need to be subtracted. Indeed, all three possible 2links and all three possible 1links are subtracted. 

Let us now evaluate this completely for the first $y$ process, so $i=1$. There is only one 1link, namely $I(x;y_1|y_2,y_3) $. There are two 2links, namely $y_1,y_2$ and $y_1,y_3$. From each of them we need to subtract the two 1links, so in total we need to subtract four 1links. Finally, there is only one 3link $y_1,y_2,y_3$, which will be decomposed as above. Hence find:
\begin{eqnarray}
 (y_1 \rightarrow x)_{total} & = & (y_1 \rightarrow x)_{1link}  + \frac{1}{2} (y_1 \rightarrow x)_{2links}
+ \frac{1}{3} (y_1 \rightarrow x)_{3links} \nonumber \\
& = & I(x;y_1|y_2,y_3)        \nonumber \\
 & + & 1/2 \left[ I(x;y_1,y_2|y_3) + I(x;y_1,y_3|,y_2) \right. \nonumber \\
& & \;\;\;\;\left. -  \left(I(x;y_1 | y_2,y_3) + I(x;y_2 | y_1,y_3)  +I(x;y_1 | y_2,y_3) + I(x;y_3 | y_1,y_2) \right)  \right]        \nonumber \\
& + &  1/3 \left[  I(x:y_1,y_2,y_3) \right. \nonumber \\
& & \;\;\;\; -   \left (I(x;y_1,y_2|y_3)+I(x;y_1,y_3|,y_2)+I(x;y_2,y_3|,y_1)  \right)  \nonumber \\
& & \;\;\;\; +  2 \left( I(x;y_1 | y_2,y_3) + I(x;y_2 | y_1,y_3)  +I(x;y_1 | y_2,y_3)  \right. \nonumber \\
& &\;\;\;\;\;  \left. \left. - I(x;y_1 | y_2,y_3) + I(x;y_2 | y_2,y_3) + I(x;y_3 | y_2,y_3)          \right) \right]
\end{eqnarray}
Because of the symmetry of the $3links$ term it is the same for all processes $y_i$.
However, both the $1link$ and the $2links$ terms are dependent on the driver process under study.
In general, for a system with $N$ drivers all links smaller than the $Nlink$ will have links that are driver-process specific.

By grouping similar terms it is possible to simplify the expression above further as:
\begin{eqnarray}
 (y_1 \rightarrow x)_{total} & = & 1/3 I(x;y_1,y_2,y_3) \nonumber \\
 &  & + 1/6 \left[ I(x;y_1,y_2|y_3) + I(x;y_1,y_3|y_2) \right] - 1/3 I(x;y_2,y_3|y_1)  \nonumber \\
& & + 1/3 I(x;y_1 | y_2,y_3) - 1/6 \left[I(x;y_2 | y_1,y_3)  +I(x;y_3 | y_1,y_2)  \right]   
\end{eqnarray}
which now defines the total contribution of $y_1$ to target $x$, decomposed in its mlink contributions.

Adding all contributions from $y_1$ to $y_3$ together we can show that indeed
\begin{equation}
I(x;y_{1:3}) = \sum_{i=1}^3 (y_i \rightarrow x)
\end{equation}
as expected. It is straightforward to extend the decomposition for $N>3$.

In terms of computational expense, the number of terms grows rapidly with the number of processes.
However, two features of the theory keep the work manageable. 
Firstly, the scheme is recursive, and secondly, the contributions from the different terms
contain many terms that are the same.
In fact, for $N=3$ we need to
calculate 3 terms of the form $I(x;y_i | y_j,y_k)$, 3 terms of the form $I(x;y_i , y_j|y_k)$
, the term $I(x;y_i,y_j,y_k,y_l)$, and $W(\hat{x})$
(or $W(\hat{x}|y_{1:N})$ but that is more expensive to calculate),
so 8 terms in total.
It is easy to show that the number of terms to be calculated is equal to
\begin{equation}
\sum_{k=0}^N \frac{N!}{k!(N-k)!} = 2^N
\end{equation}
This growth with the number of driver processes $N$ is exponential, but all mutual information calculations are independent and can be performed in parallel.

It is important to mention upfront that we do not need to calculate
probability density functions in high-dimensional spaces, but instead can use the time series directly in our calculation of the mutual informations by using the k-nearest-neighbor algorithm of \cite{Kraskov2004}.
Before we discuss how the new framework 
deals with a few well-chosen systems that illustrate its strengths and weaknesses we say a few words on how the system deals with confounders.

\section{Missing processes and Confounders} \label{sec:confouders}
Missing processes are processes that are missed when potential drivers are identified and that if included would have a strong influence on the causal strength of one or several other drivers towards the target process. An example of such a process is a confounder, which is a missed process that drives both one or more identified drivers and the target process. Let us see how these effects are represented in the framework. Assume, for ease of notation, that the system contains 3 processes, a target $x$, a known process $y$, and a missing process $z$. 

In the new framework the way to infer that a missing process is present via the causal strength of the noise term $cs(x;\eta)= W(x)/W(x|y)$ in this case. If this term is larger than expected from observational noise it is likely we missed an important driver. Practically, the effect from observational noise can be estimated by perturbing the target process series with three times the observational noise (so that the effective perturbation from the true process is twice as large) and recalculate the causal noise term, and do this several times. If the noise term in these perturbed experiments remains the same a missing process will be present. On the other hand, if the causal noise term is sensitive to these perturbations it is unlikely that an important missing process is present.

We illustrate the idea in the following simple experiment. The underlying true system is described by:
\begin{align}
x^{n+1} = & 2 y^n + z^n + \epsilon_x^n \nonumber \\
y^n = & 0.3y^{n-1} + \epsilon_y^{n-1} \nonumber \\
z^n = & 0.6z^{n-1} +  \epsilon_z^{n-1}
\end{align}
in which $\epsilon_x \sim N(0,10^{-4})$, $\epsilon_y, \epsilon_z \sim N(0,10^{-2})$. Using 20,000 time steps we calculate the causal strength of the noise without taking process $z$ into account as $cs(x;\eta)_{no\;z} = 0.301$. We add a realization of noise to the target and recalculate the causal strength, repeating this 10 times gives $cs(x;\eta)_{no\;z} = 0.308 \pm 0.005$. The difference between the causal strengths before and after the noise perturbations is about 2\%. This insensitivity to the observational noise suggests that there is a much larger contributor to the causal strength of the noise, and hence we can assume that an important driver process has been missed. 

After realizing that $z$ is an important process we recalculate the causal strength of the noise as $cs(x;\eta) = 0.086 $. Repeating the noise perturbation experiments for this case leads to $cs(x,\eta) = 0.124 $, a change of about $50\%$. The larger difference between these two suggests that no further missing processes are present.

If the missing process is a confounder it is a strong driver of both an indentified driver and the target. The now standard way to define a the presence of a confounder is via do-calculus:
\begin{equation}
p(x|do(y)) \neq p(x|y)
\end{equation}
in which $do(y)$ means that $y$ is given a certain value and all processes that influence $y$ are blocked. However, we cannot use this idea here because we are especially also interested in systems in which such an intervention would change the whole dynamics, resulting in a different system we are not interested in.

The definition used here is that $I(x^{n+1};y^{-\infty:n}|z^{-\infty:n})=0$, where $n$ is a time index and in which the conditioning is on the whole past of $z$ and we consider the causal strength of the whole past of $y$ on $x^{n+1}$. On a DAG this would mean that there is no direct arrow from $y$ to $x$ that does not go via $z$: either $y$ influences $x$ via $z$, i.e. $y^{n-1} \rightarrow z^n \rightarrow x^{n+1}$ , or $z$ is a common driver of both $y$ and $x$, i.e. $y^{n+1} \leftarrow z^n \rightarrow x^{n+1}$.  
In the first case, when $y^{n-1} \rightarrow z^n \rightarrow x^{n+1}$, we would find a nonzero causal strength between $y^{n-1}$ and $x^{n+1}$, which is correct because $y$ does drive $x$, via $z$. The framework will give the correct answer in this case. One could argue that this is not a true confounder case because of this relation.

The second case, in which $y^{n+1} \leftarrow z^n \rightarrow x^{n+1}$,  is more interesting. If we do not know about the existence of $z$ the causal strength $cs(x^{n+1};y^n)$ will be nonzero if any of the 3 processes has memory, meaning that its past is a driver of its present. This is because $x^n$ and $y^n$ are driven by $z^{n-1}$, and memory in either $x$, $y$, or $z$ will result in a connection between $y^n$ and $x^{n+1}$. Without having knowledge of $z$, this is a correct answer.

The question is what happens when $z$ comes to our attention. Remember that the confounding nature of $z$ on the $(x,y)$ relation is defined as $I(x^{n+1};y^{-\infty:n}|z^{-\infty:n})=0$. In this case the total causal strength of the whole past of $y$ would be:

\begin{eqnarray}
cs(x^{n+1};y^{-\infty:n}) & = & \frac{I(x^{n+1};y^{-\infty:n}|z^{-\infty:n})}{W(x^{n+1}|y^{-\infty:n},z^{-\infty:n})}  +  \frac{1}{2} \frac{\left(I(x^{n+1};y^{-\infty:n}) - I(x^{n+1};y^{-\infty:n}|z^{-\infty:n}) \right)}{W(x^{n+1}|y^{-\infty:n},z^{-\infty:n})}\nonumber \\
& = & \frac{1}{2} \frac{I(x^{n+1};y^{-\infty:n})}{W(x^{n+1}|y^{-\infty:n},z^{-\infty:n})}
\label{eq:confounders2}
\end{eqnarray}

Hence, this nonzero causal strength shows that the framework suggests that $y$ does influence $x$, while in fact $z$ is driving both $y$ and $x$ separately. However, if $z$ is driving $y$, than $y$ has information from $z$, so $y$ has information that drives $x$. The framework does recognize the information $y$ has on driving $x$. Since the framework does not only calculate causal strength but also all mlink contributions including $I(x^{n+1};y^{-\infty:n}|z^{-\infty:n})=0$, it recognizes the confounding nature of $z$.  

This short discussion demonstrates how missing processes are identified and how subsequently confounders are identified in the causal framework. We have a way of estimating the influence on the target of missing processes relative to the observational noise in the system, allowing for identification of their presence, and we identify true confounders via 1links.

\section{Comparison with other frameworks}
Since our framework is specifically developed with nonlinear systems in mind, we only compare to other methods that allow for nonlinear interactions. In table I a comparison of capabilities is provided. As mentioned in the introduction, Transfer Entropy as introduced by  \cite{Schreiber2000} and further developed for high-dimensional systems by e.g. \cite{Runge2018} does only a partial decomposition, similar to a 1link in the new framework. The PID framework has no unique definition of Unique, Synergistic, and Redundant information. As we will see in the first example in the next section, the simplest 3-variable system already leads to interpretation problems that suggest this decomposition is not that useful. Only the new framework provides a full decomposition of the causal strength over its different contributions, has a unique normalization that does not depend on application, is complete in the sense that it allows for confounder detection, and generates complete causal webs that are much richer than DAGs because they allow for merging of directed edges.
For two different causal web styles, see figures 2 and 8.

Our framework has some parallels with the framework that Runge developes in \cite{Runge2015a}. 
His framework aims to answer the question how strong the indirect causal influence is of a process on a target process, where the direct causal influence is defined via a transfer entropy. Specifically, the paper concentrates on the specific influence of a process $y$ that is a few time steps in the past of the target process $x$, and where $y$ influences other processes $z$ that in their turn influence $x$. The interaction information from $y$ via $z$ is defined as the mutual information of all paths between $y$ and $x$ minus the mutual information of all paths between $y$ and $x$ conditioned on process $z$.
The paper restricts the analysis to causal systems that can be represented by a DAG, while our framework is more general than that because we explicitly take nonlinear interactions between processes into account which cannot be represented on a such a graph.

\begin{table}
\caption{Comparison of causal discovery frameworks.}
\centering
\footnotesize
\begin{ruledtabular}
\begin{tabular}{lllll}

\textbf{Characteristic}	& \textbf{New framework}	& \textbf{Transfer Entropy} &  \textbf{PID} & \textbf{CCM}	\\
\hline
Total causal strength   & $cs(x,y)$ 	&  1links $I(x;y|z)$    	&    $U+S+R$ per cause &  - \\
Decomposition & 1link, all 2links, 
&  1link per cause		&  $U,S,R$ per cause & binary target-driver, 	\\
  & ..., Nlink per cause 	&  & & no conditioning \\
Normalization  & with $W(x|y,z)$ 			& application dependent 		&    - 	& - \\
Missing process detection    & via $cs(x;\eta)$ & -  	&    -  & -  \\
Confounder detection    & via $cs(x;\eta), $ & zero 1links  	&    -  & -  \\
   & followed by zero 1links  &   	&      &   \\
Graphical representation     & full causal web,  & DAG  	&    hypergraph ?  & DAG  \\
     & (see e.g. fig8) hypergraph &   	&      &   \\
\end{tabular}
\end{ruledtabular}
\end{table}

\section{Examples}

Several examples are discussed to illustrate the behavior of the new framework.
We start with linear models with Gaussian noise, then discuss nonlinear models 
without interactions between the terms, followed by models with nonlinear interactions
and finally the Lorenz 1963 model. 

All information theoretic quantities were calculated using the k-nearest-neighbor algorithm of \cite{Kraskov2004}, where the number of nearest neighbors is set to $4$, with little sensitivity to the actual number. To increase the numerical accuracy the target and all drivers are transformed via the CDF of a Gaussian to a truncated Gaussian, specifically cutting 4.25\% of the full distribution off each tail. The reference density for the certainty calculation is the  Lorentz-Cauchy density. In section 6.3 the influence of different reference densities is investigated in detail.

\subsection{Memory-limited models}

The following models are special in that their temporal memory is strongly limited, allowing us to concentrate on local-in-time relations.
Furthermore, the models are simple enough so that they can be represented on a standard graph, except for model 6.
Table \ref{tab:model-eqs-1-3} shows the first 3 models that we used to generate time series, on which we then test the causal discovery framework.
We generated 100 time series from each model of length 50,000 steps and calculated the mutual informations and 
conditional mutual informations as needed. The results of the experiments are presented in Table \ref{tab:results-models1-3}. 

\begin{table}
\caption{Underlying model equations, and characteristics of the noise terms.}
\label{tab:model-eqs-1-3}
\centering
\begin{ruledtabular}
\begin{tabular}{llll}

\textbf{Model}	& \textbf{x}	& \textbf{y} &  \textbf{z}	\\
\hline
model  1   & $x^{n+1} = 2y^n + z^n + N(0,10^{-4})$ 	&  $y^n = N(0,1)$    	&    $z^n = N(0,1)$   		  \\
model 2    & $x^{n+1} = z^n + N(0,10^{-4})$ 			& $y^{n-1} = N(0,1)$  		&    $z^n =y^{n-1}  + N(0,1)$ 	\\
model 3    & $x^{n+1} = z^n + N(0,10^{-4})$ 			& $y^{n+1} = z^n + N(0,10^{-2})$  		&    $z^n =N(0,1)$ 	\\
\end{tabular}
\end{ruledtabular}
\end{table}

Model 1 is perhaps the most simple model one can think of, and can be represented on a graph as $y \rightarrow x \leftarrow z$. It is linear and has no memory, so interpretation of the terms should be straightforward. The conditional mutual informations, the 1links, are larger than the mutual informations between $y$ and $x$ and between $z$ and $x$.
This means that the interaction information is negative, and the reason is that without conditioning the variable $z$ acts as noise in the mutual information calculation of $y$ and $x$, and similarly for $y$. The causal strength of $y$ to $x$ is 1.6 times larger than that of $z$ to $x$ (0.56/0.35),
with a small contribution for the noise process. If only the 1links would be taken into account, the ratio of the $y$ contribution to the $z$ contribution is much lower, 1.1, due to the omission of the 2link contributions. In this simple model the conditional mutual information is equal to the transfer entropy,  and it is interesting to see how transfer entropy suggests very similar contributions from $y$ and $z$, while the causal strength of the former is expected to be much larger considering the actual model, as correctly indicated by our framework.

It is also interesting to connect these results to the PID framework. The form of model 1 suggests that there is no synergy and no redundancy in this system because $y$ and $z$ are completely independent when driving $x$. However, $I(x;y|z)>I(x;y)$ and hence $R(x;y,z)-S(x;y,z)<0$ in the PID framework, so that the synergy has to be nonzero, as all contributions are non-negative in the PID framework. The only way to keep consistency in the PID framework is to introduce the influence of unaccounted-for processes ('noise'), as we do in our discussion above on the 2links, but PID does not include 'noise' terms. The words 'unique', 'synergy', and 'redundancy' are difficult to define even in this simple system. If we make the straightforward choice $U(x;y|z) = I(x;y|z)$ we find, see equation (\ref{eq:UplusS}) in the introduction, that $S(x;y,z)=0$, which leads to  $R(x;y,z)<0$ which is inconsistent with the non-negativeness of the variables in the framework. If, instead, we choose $R(x;y,z)=0$ the unique contribution of $y$ on $x$ becomes the mutual information $I(x;y)$, while we know that this mutual information also contains influence from $z$.

\begin{table}
\caption{(Conditional) mutual informations and total information flows. Typical uncertainties are $0.005$, based on 10 random realizations of the time series.}
\label{tab:results-models1-3}
\centering

\begin{ruledtabular}
\begin{tabular}{cccc}
\textbf{Estimate}   			& Model 1         & Model 2        	& Model 3 	\\
\hline
I(x;y|z)			        		& 2.99		& 0.00 		&0.00			\\
I(x;z|y)		         		& 2.30 		& 2.33 		&2.33		\\
I(x;y)			         		& 0.80 		& 0.33 		&0.00			\\
I(x;z)       			       	 	& 0.11 		& 2.66		&2.33		\\
$I(x;y;z)  $					& -2.19 		& 0.33		&0.00	\\
I(y;z)  	          			& 0.00          	& 1.42	        &2.65			\\
$(y \rightarrow x)_{total}$ 		& 1.91 		& 0.17		&0.00		\\
$(z \rightarrow x)_{total}$ 		& 1.19	        & 2.50		&2.33			\\
$W(\hat{x})$ 				& 0.31		& 0.30		&0.29		\\
$W(\hat{x}|y,z)$ 			& 3.41 		& 2.96		&2.61			\\
$cs(x;y)$					& 0.56 		& 0.06	        &0.00			\\
$cs(x;z)$  					& 0.35 		& 0.84		&0.89			\\
$cs(x;\eta)$				& 0.09 		& 0.10		& 0.11			\\

\end{tabular}
\end{ruledtabular}
\end{table}

Model 2 is a system in which $z$ acts as a gateway for the information flow between $y$ and $x$, graphically $y \rightarrow z \rightarrow x$. 
The only nonzero contributions in model 2 are those between $x^{n+1}$, $z^n$ and $y^{n-1}$ since the system variables have no memory.
This leads to positive interaction information because conditioning on $z$ in $I(x;y|z)$ destroys the connection between $y$ and $x$. 
The mutual information between $y$ and $x$ is nonzero without this conditioning, showing that there is 
information flow from $y$ to $x$ in this system because knowing $y$ does provide information about $z$, and hence information about $x$; the 1links are just unable to pick this up.
As expected, the causal strength of $z$ to $x$ is much higher than that of $y$ to $x$. 
The reason for this small $y$ contribution is that the noise $\eta_z$ is of the
same order of the signal $y$, making the $I(x;y)$ dominated by noise. This can be seen clearly when we substitute the expression for $z^n$ in 
model 2: $x^{n+1} = y^{n-1}+ \eta_z^n + \eta_x^{n+1}$. Indeed, lowering the noise in $z$ does make $I(x;y)$ much larger
(not shown). It is also interesting that if we remove $y$ from the causal calculations, so the underlying model remains model 2, but we only consider $x$ and $z$, the total causal strength from $z$ to $x$ increases to $0.90$. Hence in this case $z$ takes up the causal strength of $y$, which is exactly what the framework should do.

Model 3 can be represented as $y \leftarrow z \rightarrow x$, so, $z$ drives both $y$ and $x$. The measures are calculated between $x^{n+1}$ and $y^n$ and $z^n$ as the drivers need to be lagged at least one time unit from the target for causal influence.  We see that any measure between $y$ and $x$ is zero, and indeed the causal strength between $y$ and $x$ is zero, as it should be. The difference between the case discussed in the confounder section is that the processes have no memory in this model. 
 
It is interesting to discuss what we could infer from Transfer Entropy, which only works with the first two rows in Table \ref{tab:results-models1-3}. It can distinguish between models 1 and 2, but not between models 2 and 3. Furthermore, in model 2 it does not see that $y$ influences $x$ via $z$. Finally, it has no way to infer if strong confounders are present or not.

We now study how the framework reacts to time series with memory. In Model 4 in Table \ref{tab:model-eqs-4-6} the evolution of $z$ is not influenced by $x$ and $y$, but $z$ is a driver for both $x$ and $y$, graphically $x \leftarrow z \rightarrow y$. Table \ref{tab:results-models4-6} shows the results for various quantities from the new framework. 

\begin{table}
\caption{Underlying model equations, and characteristics of the noise terms.}
\label{tab:model-eqs-4-6}
\centering
\begin{ruledtabular}
\begin{tabular}{llll}
\textbf{Model}	& 	& & 	\\
4    & $x^{n+1} = 0.4x^n +0.4z^n + N(0,10^{-4})$ 			& $y^{n} = 0.5y^{n-1}+0.5z^{n-1} +N(0,10^{-2})$  		&    $z^{n} =0.4z^{n-1}+N(0,10^{-2})$ 	\\
5    & $x^{n+1} = 0.6x^n +y^nz^n+0.3z^n + N(0,10^{-6})$ 			& $y^{n} = 0.3y^{n-1}+N(0,10^{-4})$  		&    $z^{n} =y^{n-1}+N(0,10^{-4})$ 	\\
6    & $x^{n+1} = w^n + 0.6 y^{n-1} + 0.4 z^{n-1} + N(0,10^{-4})$ 			& $y^{n} = N(0,1) \;\;\;\;\;\;\;\;$         $z^n = N(0,1)$ & $w^{n} = y^{n-1} + 4 z^{n-1} + N(0,1)$	\\
\end{tabular}
\end{ruledtabular}
\end{table}

The conditional mutual information of $y$ and target $x$, given the past of $x$ and $z$ is zero, which, in a directed graph would mean that $y$ is not a driver of $x$. The causal strength $cs(x;y)$ is nonzero, however. This reflects the possibility that $y$ influences $x$ jointly with other drivers. It turns out that the memory in the variables is important in understanding what happens. In Table \ref{tab:results-models4-6} variable $y$ denotes the combination $y^n,y^{n-1}$, approximately the whole process $y$ in the past of $x^{n+1}$. Similarly $z$ denotes $z^n,z^{n-1}$. Model 4 shows that if $y=(y^n,y^{n-1})$ is known, then we know also about $z^{n-1}$. But $z^{n-1}$ drives $z^n$, which in turn drives $x^{n+1}$. The framework does pick up this link, but should it? In fact, it should, as we can write the evolution equation of $x$ purely in terms of $y$, as:
\begin{eqnarray}
    x^{n+1} & = &  0.4x^n +0.4z^n + N(0,10^{-4}) \nonumber \\
    & = &  0.4x^n +0.16z^{n-1} + N(0,10^{-2}) \nonumber \\
    & = & 0.4x^n + 0.32y^n - 0.16y^{n-1}+ N(0,2x10^{-2})
\end{eqnarray}
Hence, $y$ is a true driver of $x$.
The main driver of target $x$ is $z$ followed by the past of $x$. This ordering makes sense as $z$ also influences the past of $x$, while $x$ does not influence $z$. While $y$ can be considered a true driver of $x$ via the equation above, the noise term in that equation is much larger than in the original equation, explaining its small causal strength.

\begin{table}
\caption{(Conditional) mutual informations and causal strengths for Model 4,5 and 6. Typical uncertainties are 0.005, based on 10 random realizations of the time series. For Model 5 and 6 the variable $x_d$ is the variable $x$ in the past of target $x$. All drivers are total contributions from one and two steps lagged behind the target.}
\label{tab:results-models4-6}
\centering

\begin{ruledtabular}

\begin{tabular}{cccc}
\textbf{Estimate}				& Model 4 & Model 5 & Model 6		 \\
\hline
$I(x;y|x_d,z)$ or $I(x;y|z,w)$		&0.00	&0.00	& 0.07		\\
$I(x;z|x_d,y)$ or $I(x;z|y,w)$		&1.18	&0.36	& 0.00 			\\
$I(x;x_d|y,z)$ or $I(x;w|y,z)$		& 0.15	&0.22	&0.00 			\\
$cs(x;y)$ 	         				&0.05	&0.17	& 0.20 		\\
$cs(x;z)$   					&0.65	&0.39	& 0.29 		\\
$cs(x;x_d)$ or $cs(x;w)$    		&0.14	&0.22	& 0.44 			\\
$cs(x;\eta)$ 					&0.16	&0.22	& 0.07 		\\

\end{tabular}
\end{ruledtabular}
\end{table}

Model 5 shows an example in which target $x$ is driven by its past, by $y$, and by $z$, while $z$ is completely driven by $y$. So $y$ and the past of $x$ are the driving processes in this model. However, $I(x;y|x_d,z)=0$, while $I(x;z|x_d,y)=0.36$. This cannot happen in a model that can be represented on a Directed Acyclic Graph, but Model 5 has nonlinear interactions between its drivers, so cannot be represented so. Any method that is based on a DAG representation will miss the importance of $y$ for driving $x$. This shows the importance of including the higher-order links in the causal network. 

One can argue that Model 4 and 5 are qualitatively the same, based on Table \ref{tab:results-models4-6}. Can the framework distinguish between these two totally different underlying systems? In fact it can, but one has to dive deeper into the decomposition. In the nonlinear Model 5 we find (not shown) that the 2link $\hat{I}(x^{n+2};z^{n+1},y^n|rest) = 0.21$ (see Eq. (\ref{eq:hat}) for its definition), while this contribution is $0$ for the linear model. This is the 2link between $z^{n+1}$ and $y^n$ when they drive $x^{n+2}$, where we subtracted the 1links to avoid double counting, see section 4. In the linear Model 4 this interaction is already covered via the 1link of $z$ with target $x$, so the 2link does not provide much more information. However, in the nonlinear Model 5 there is crucial information added by the interaction of $y$ and $z$ over just the 1links. Furthermore, the fact that $y$ is lagging behind $z$ shows that $y$ is driving $z$. Note that $\hat{I}(x^{n+2};y^{n+1},z^n|rest)$ is zero in both models, so it does not show that $z$ is driving $y$ in Model 5. The reason is that in the linear model that interaction is taken up via the 1links. 

In Model 6 we show that even in a linear model counterintuitive things can happen. Also in this model $y$ stands for $(y^n,y^{n-1})$, etc.  In all earlier model examples the order of importance of $y$ and $z$
on $x$ doesn't change using just the 1links or the full causal strengths from the complete framework. As mentioned earlier, this is indeed the case 
for any 3-variable model because the interaction information is symmetric in $y$ and $z$. 
However, when more variables are introduced this ordering can change, as shown in the last column of Table \ref{tab:results-models4-6}.
In terms of 1links (conditional information) one would expect that
$y$ is most important for $x$, and $z$ and $w$ are not important at all.
However, taking all links properly into account we find that $w$ is most important, then $z$ and then $y$.  
So roles have completely reversed. The reason for the reversal of $y$ compared to $z$ is that as soon as we do not condition on $w$, $z$ is more important for $x$ than $y$.
Similarly, $w$ changes roles with $z$ because conditioning on $y$ and $z$ makes $w$ just a noise process for $x$, so the 1link has zero conditional mutual information. Transfer Entropy, which only uses the 1st three rows of Table \ref{tab:results-models4-6}, would misinterpret the order of importance of the driving processes. 
\subsection{The Lorenz 1963 model}
We now apply the framework to the well-known Lorenz 1963 model, with model equations:
\begin{eqnarray}
\frac{dx}{dt} & =  & \sigma(y-x) \nonumber \\
\frac{dy}{dt} & = &  \rho x -xz -y \nonumber \\
\frac{dz}{dt} & = & xy - \beta z 
\label{eq:L63eqs}
\end{eqnarray}
The interesting aspect of this system of equations is that it cannot be fully represented on a standard graph.
We generated time series of $x$, $y$, and $z$ for 50,000 time steps using a Runga-Kutta 4 scheme with time step 0.01, starting very close to the attractor at (1.50887, -1.531271, 25.46091).
We use as drivers the 3 processes $x$, $y$, and $z$, and as target process the time series of $x$ shifted forward one time step.
To make this a realistic experiment we added Gaussian noise of variance 0.01 to each time series after integrating the Lorenz 1963 equations, i.e. adding observational noise.
We are trying to reconstruct the causal structure of the system using only its noisy time series.

\begin{figure}
\centering
\includegraphics[width=0.8\linewidth]{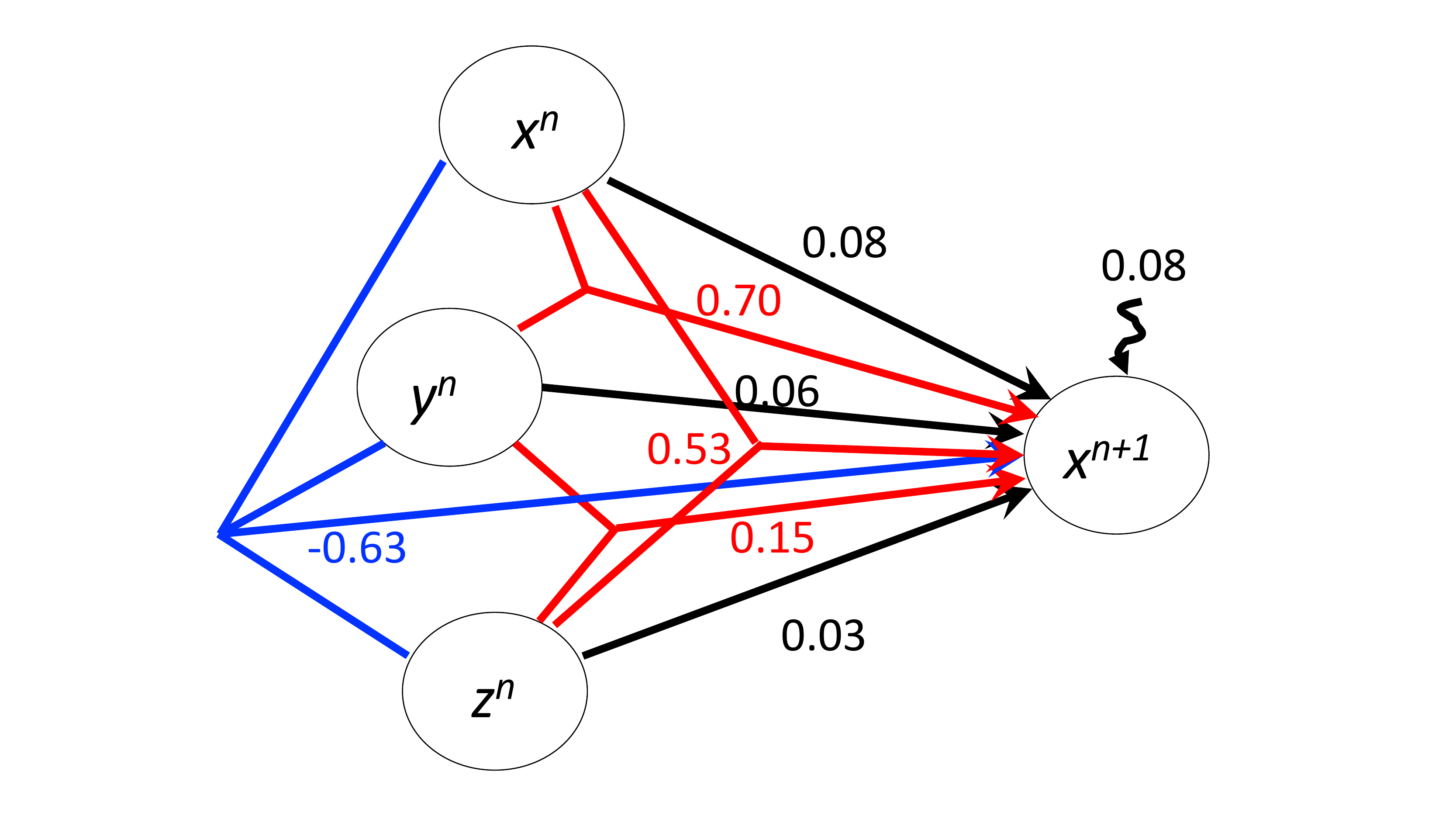}
\caption{Causal connections for the Lorenz 1963 system  between driver processes $x^n$, $y^n$ and $z^n$, and target process $x^{n+1}$, where $n$ is the time index. 
The black arrows denote the direct connections between drivers and target, the 1links. 
The red arrows show the 2links between 2 drivers and the target, and the blue lines denote the 3link. All values represent causal strengths and have been normalized by the total certainty $W(x^{n+1}|x^n,y^n,z^n)$.}
\end{figure}   

Figure 2 shows the causal strength of the links for the $x$ target. 
Perhaps surprisingly initially, the 1link contributions (represented by the black arrows) are all smaller than some of the 2links. 
Looking at the equation for $x$, the small size of $\Delta t$ would suggest that to a very good approximation
$x^{n+1} = x^n + \epsilon^n$, where $|\epsilon^n| << |x^n|$.
However, unlike correlations, the actual size of the variables is not important; rather the narrowness of the joint probability density functions determines the size of the causal strengths.
This is immediately clear when it is realized that a mutual information value is independent of a single-variable nonlinear 
monotonic transformation.
For the Lorenz 1963 system, if we know $x^n$, $x^{n+1}$ can be larger or smaller, that depends 
on if we are on the upward or the downward branch of a Lorenz wing. 
However, knowing $x^n$ and $y^n$ tells us in which branch of a wing the system is, and hence we know quite well if $x^{n+1}$
will be larger or smaller than $x^n$. 
Hence knowing $x^n$ {\em and} $y^n$ is much more valuable for predicting the value of $x^{n+1}$ than $x^n$ alone,
and indeed the 2link is about a factor 9 larger (0.70) than the 1link $I(x^{n+1};x^n|y^n,z^n)$ (0.08). 
But there is more to this.

Figure 2 shows that $x^n$ and $z^n$ have a strong causal relation with $x^{n+1}$, of value 0.53,
while $z^n$ does not even appear in the governing equation for $x^{n+1}$. We can learn a lot from this. 
Firstly, the framework is not optimized to find the physical laws that govern the underlying dynamics.
This is not surprising as, as mentioned above, mutual informations cannot distinguish between nonlinear and linear relations,
in the sense that they are insensitive to a single-variable nonlinear monotonic transformation.
However, we now see that it cannot even determine from the 2links if a variable is present in one of the governing equations of a system.
This means that information has to flow in from what happens before time $n$, so from the larger scale dynamics. (Note that this connection is not due to the use of a RK4 numerical scheme as it is also present when using an Euler scheme.)
At the larger scale dynamics, knowing $x^n$ and $z^n$ does tell us the wing and the direction of flow, so it is known if
 $x^{n+1}$ will be larger or smaller than $x^n$: the direction of flow is known.

This idea is strengthened by the fact that the 2link from $y^n$ and $z^n$ to $x^{n+1}$ is smaller, 0.15.
This lower value is related to the fact that in the $y-z$ plane the two wings overlap to a large extent, and it is difficult
to know which wing is which, and hence what the value of $x^n$ is. Thus it will be difficult to predict $x^{n+1}$.

Finally, the 3link is negative and quite large. The 3link contains that flow of information from all drivers towards the target $x$ 
after the 1link and 2link contributions have been subtracted. 
From this we can understand that its negative value indicates that the 2links and 1links contain redundant information,
for instance, the 2links $x,y$ and $x,z$ contain overlapping information that needs compensation.

To find the total contribution of $x^n$ from Figure 2 we take the 1link, and 1/2 times the 2links it is involved in,
and 1/3 of the 3link it is involved in, leading to
$0.08 + (1/2)(0.70+0.53) + (1/3) (-0.63) = 0.48 $.
Using this methodology, we find for the total contributions of $y$ and $z$ $0.27$ and $0.15$, respectively, leaving $0.09$ for 
the noise contribution, as detailed in Table \ref{L63CStable}. This table does suggest that $z$ is less important than $x$ and $y$ for $x^{n+1}$, but its contribution is not zero.

\begin{table}\label{L63CStablexyz}
\caption{Causal strength for Lorenz 1963 model, 1 time lag, with standard deviations of 0.005\% }
\centering
\begin{ruledtabular}
\begin{tabular}{cccccc}

Estimate			& value & Estimate			& value & Estimate			& value 	 \\
\hline
$cs(x^{n+1},x^n)$		& 0.485 & $cs(y^{n+1},x^n)$		& 0.260 	 &	$cs(z^{n+1},x^n)$		& 0.173 		\\
$cs(x^{n+1},y^n)$  		& 0.274 & $cs(y^{n+1},y^n)$  		& 0.545 	 &	$cs(z^{n+1},y^n)$		& 0.130 		\\
$cs(x^{n+1},z^n)$  		& 0.151 & $cs(y^{n+1},z^n)$  		& 0.135 	 &	$cs(z^{n+1},z^n)$		& 0.584 		\\
$cs(x^{n+1},\eta^n)$		& 0.090 & $cs(y^{n+1},\eta^n)$		& 0.062	 &	$cs(z^{n+1},\eta^n)$		& 0.114 	\\

\end{tabular}
\end{ruledtabular}
\label{L63CStable}
\end{table}

Figures 3 and 4 show similar diagrams for the $y$ and $z$ targets. 
The first thing that catches the eye is that 2links containing the target 1 step back in time are again large.
Also here the 3link cannot be neglected and is negative for both the $y$ and the $z$ target. This means that the 1- and 2links  contain redundant information that needs compensation, similar to what we found for the $x$ target. For the $z$ variable as target, the 1link with $z$ one step back in time is much larger than for the targets $x$ and $y$. The main reason for this is that $z$ is independent on the wing the system is in. The product of $x$ and $y$ tells whether $z$ is increasing or decreasing.
 
\begin{figure}
 \includegraphics[width=0.8\linewidth]{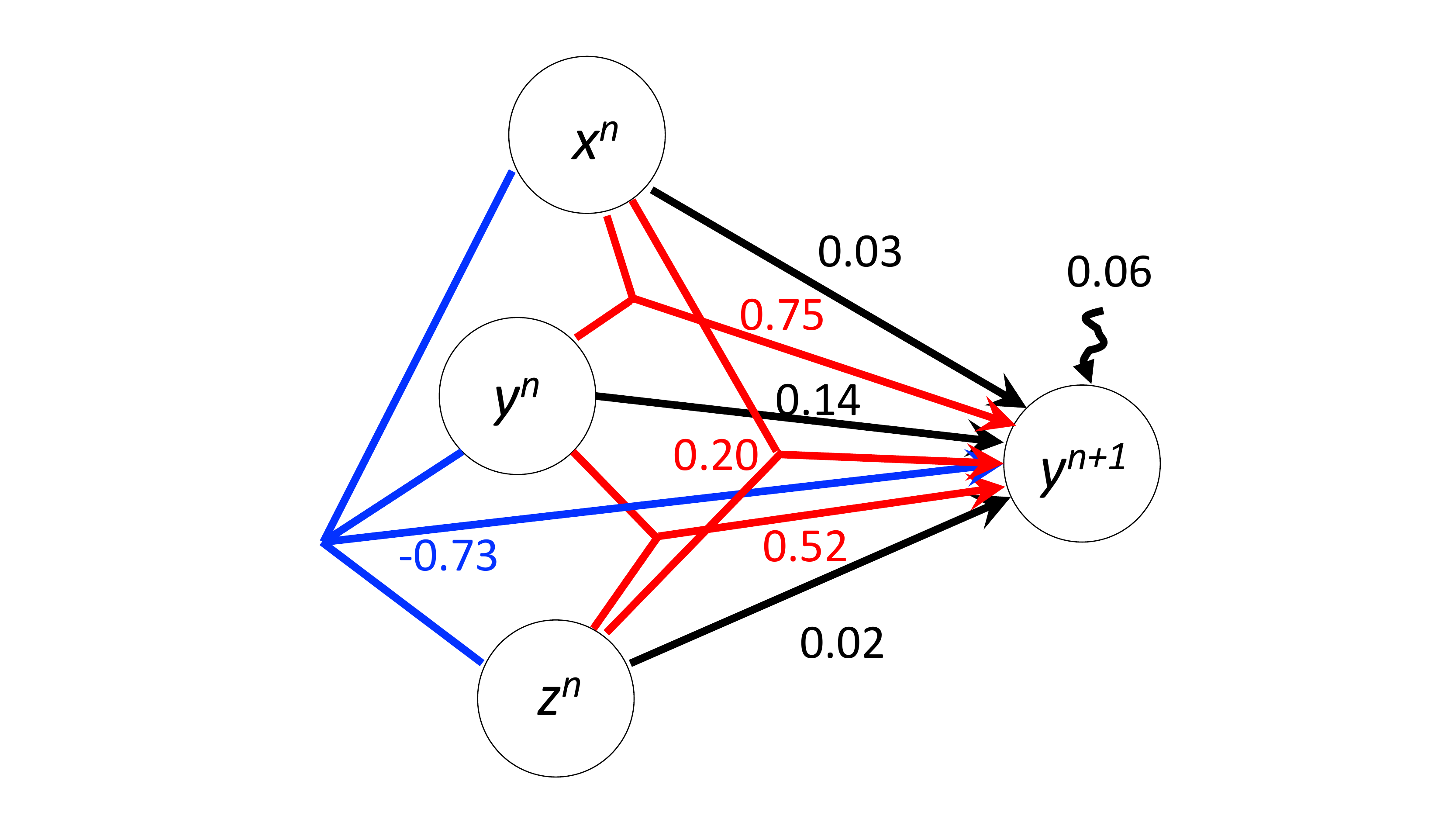}
 \caption{Causal connections between driver processes $x^n$, $y^n$ and $z^n$, and target process $y^{n+1}$. 
The black arrows denote the direct connections between drivers and target, the 1links. 
The red arrows show the 2links between 2 drivers and the target, and the blue lines denote the 3link}
 \end{figure}
 \begin{figure}
\includegraphics[width=0.8\linewidth]{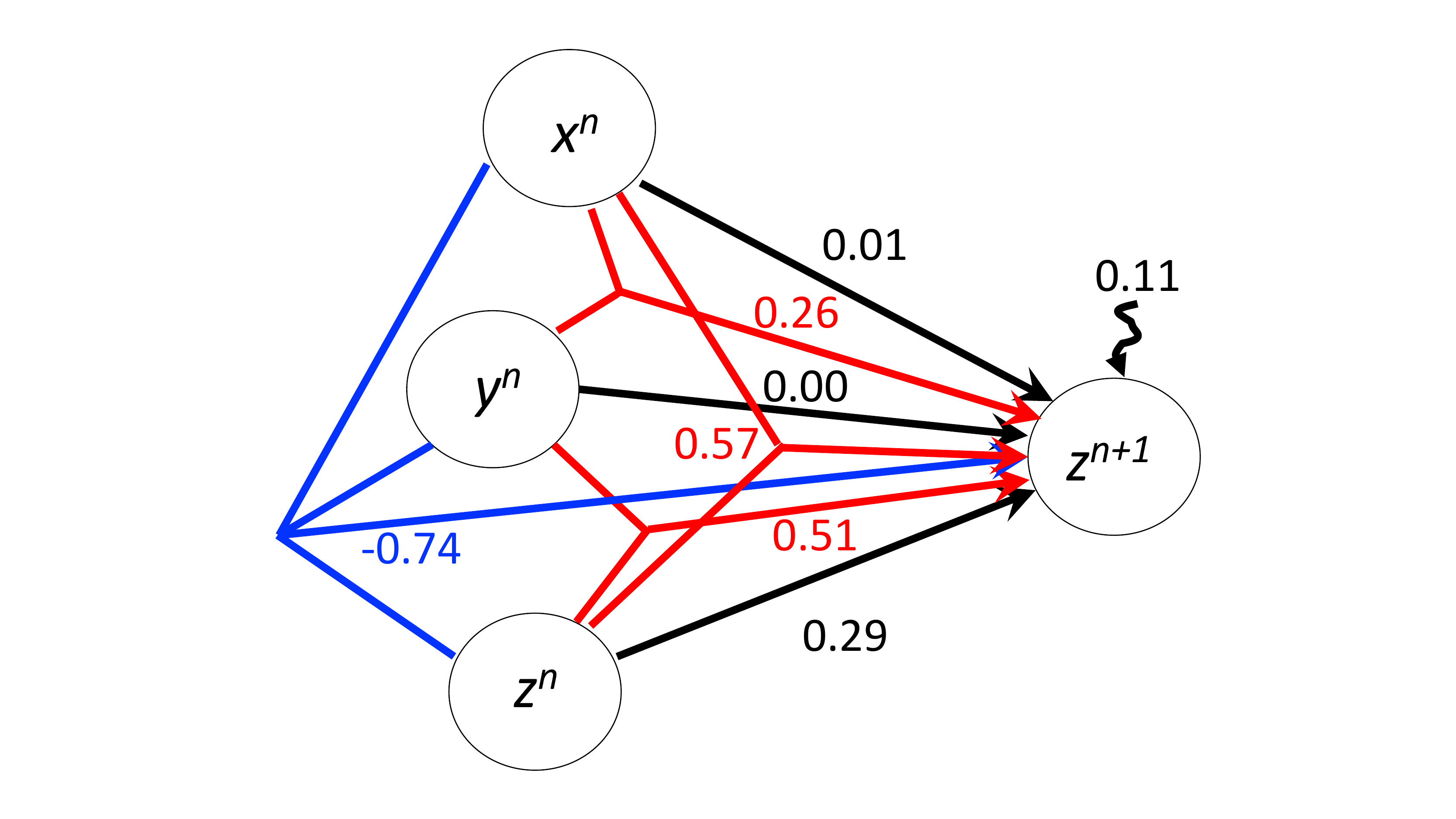}
\caption{The same for target process $z^{n+1}$. All values have been normalized by the total certainty $W(y^{n+1}|x^n,y^n,z^n)$ and $W(z^{n+1}|x^n,y^n,z^n)$, respectively.}
\end{figure}

We can again calculate the causal strengths of each variable $x^n,y^n,z^n$ to $y^{n+1}$ and similarly for $z^{n+1}$ and the results are depicted in Table \ref{L63CStable}. Given the underlying evolution equation, it is not surprising that $x$ is more important than $z$ for $y$.

We see from the causal strengths that they are much closer to the governing equations than e.g. the 1link contributions.
On the other hand, the 1link and 2link contributions seem to tell us more about the underlying large-scale structure.
This is a quite interesting feature of the new framework that we will elaborate on in a further study.

\subsection{The coupled Lorenz 1963 model}

As a final example of the workings of the framework we study 2 Lorenz 1963 systems in which one forces the other. This system has also been studied by \cite{Staniek2008} and reads:
\begin{align}
\frac{dx_1}{dt} & =   \sigma(y_1-x_1) + \epsilon(x_2-x_1) &\frac{dx_2}{dt} & =   \sigma(y_2-x_2) \nonumber \\
\frac{dy_1}{dt} & =   \rho x_1 -x_1z_1 -y_1 &\frac{dy_2}{dt} & =   \rho x_2 -x_2z_2 -y_2 \nonumber \\
\frac{dz_1}{dt} & =  x_1y_1 - \beta z_1 & \frac{dz_2}{dt} & =  x_2y_2 - \beta z_2  
\label{eq:L63eqscoupled}
\end{align}
The coupling strength $\epsilon$ is varied from $0$ to $9$, the latter corresponding to complete synchronization of the two systems. The coupled system was discritized with a Runge-Kutta 4 scheme with a time step of 0.01. A spin up run of $10^4$ time steps was performed before each experiment, each of which lasted $50,000$ time steps. We show averaged results based on 10 runs starting with different initial conditions. 

 \begin{figure}
\includegraphics[width=0.7\linewidth]{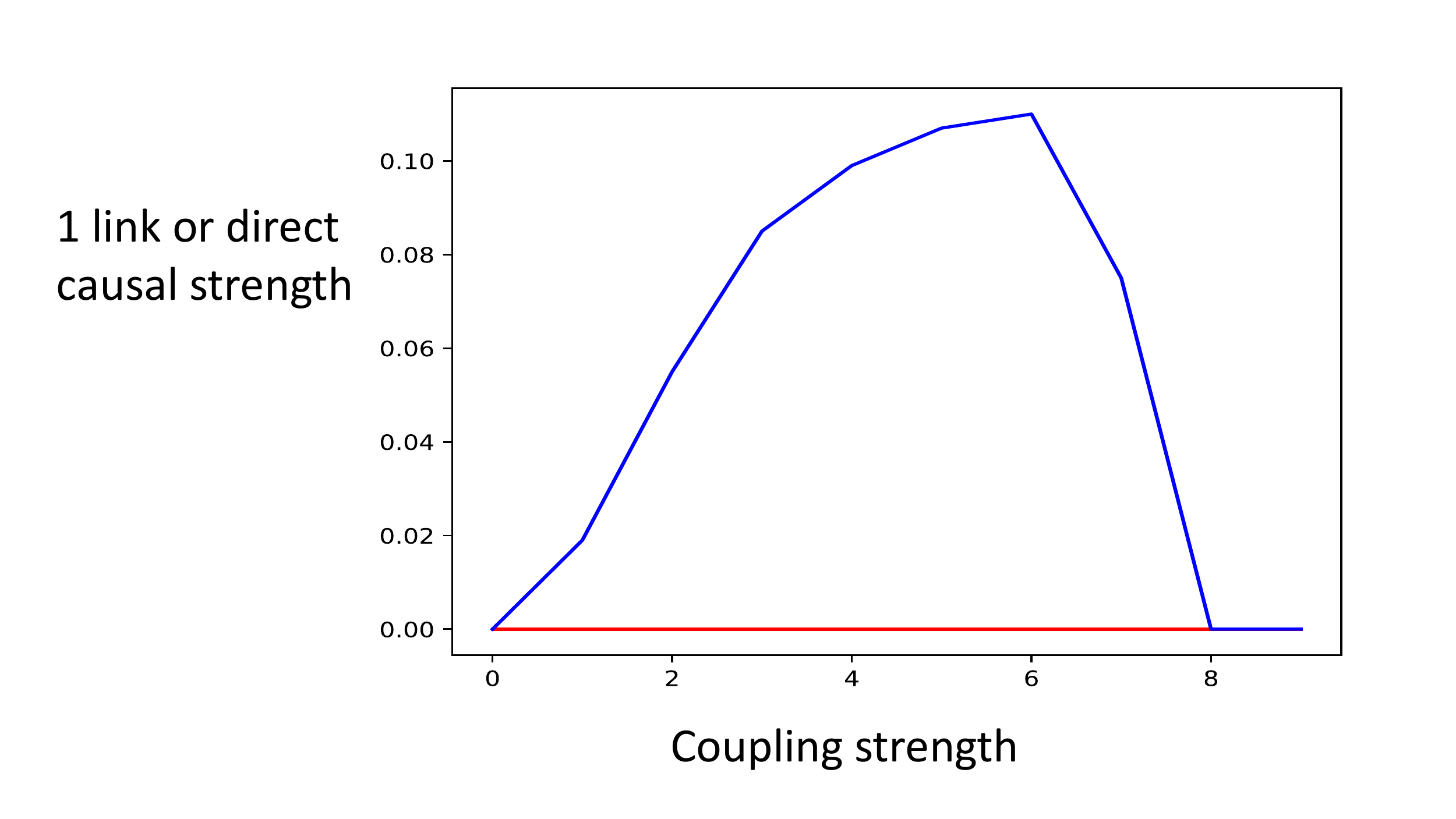}
\caption{Direct causal strength, or 1link, from $x_2$ to $x_1$ (blue) and from $x_1$ to $x_2$ (red) for the coupled Lorenz 1963 system as function of the coupling strength $\epsilon$.}
\end{figure}

Figure 5 shows the direct causal strength or the 1link of $x_1$ on $x_2$ with the red line, and of $x_2$ on $x_1$ with the blue line. We see that $x_1$ does not drive $x_2$ at any coupling strength, consistent with the model equations. The driving of $x_2$ towards $x_1$ shows a maximum at $\epsilon=6$ and drops to zero for higher coupling strengths. The reason for this is that for larger coupling strengths $x_1$ and $x_2$ are close to full synchronization, and hence the coupling term $\epsilon(x_2-x_1)$ becomes smaller and smaller. If the underlying structure of the system is not known (and it is assumed unknown in our experiment) and the coupling strength is large, say $8$ or higher, one could draw the conclusion that the two systems are uncoupled by just looking at these 1links. This shows that the 1links, e.g. such as provided by Transfer Entropy and its variants \cite{Staniek2008}, do not provide enough information.

 \begin{figure}
\includegraphics[width=0.7\linewidth]{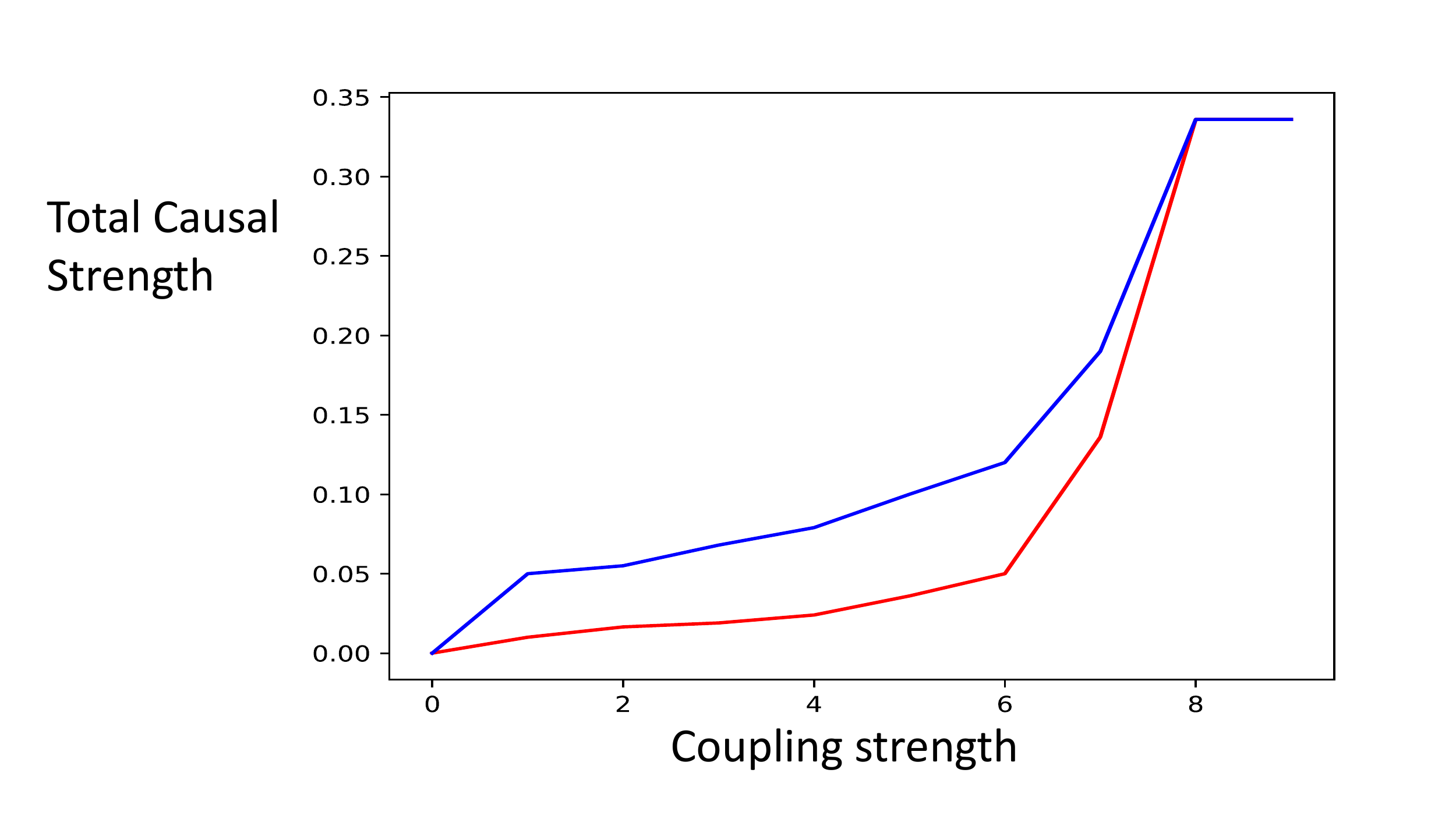}
\caption{Total causal strength from $x_2$ to $x_1$ (blue) and from $x_1$ to $x_2$ (red) for the coupled Lorenz 1963 system as function of the coupling strength $\epsilon$.}
\end{figure}

Instead, looking at the total causal strengths in figure 6, we see that $x_2$ is causing variations in $x_1$ (blue line), slowly growing with the coupling strength, with faster growth after $\epsilon=6$. As we have seen in figure 5, this is when synchronization sets in, and $x_1$ and $x_2$ become very similar. The total causal strength from $x_1$ to $x_2$ is much lower. It is non-zero because the two systems start to synchronize slightly for low coupling strengths, so that $x_2$ has some information on $x_1$. Only when the coupling strength is larger than $6$ synchronization becomes so strong that the two curves start to behave quite similar. 

\subsection{Sensitivity to reference density}
We close this section of examples by studying the sensitivity of the results to the reference density. The reference density only shows up in the self certainty and the total certainty, not in the mutual informations. Hence it will not influence the causal strength of the driver processes relative to each other. However, it will influence the size of the self certainty, and through that the relative size of the 'noise' term compared to the other drivers. 

In all results above we used the Lorentz-Cauchy density, with mean equal to the target sample mean, and width parameter $\gamma = \sqrt{(e/8 \pi)} \sigma_x$, such that the entropy of the density is equal to that of a Gaussian with standard deviation $\sigma_x$, but with infinite variance. We will compare these results with those from two other reference densities, the Gaussian density with mean and variance equal to that of the target, and a density that is uniform on the interval spanned by the range of the target samples $[\min(x),\max(x)]$ and zero outside that range. The Gaussian and the uniform density are extreme cases in the sense that if the target is Gaussian distributed the noise contribution will be zero, while a uniform density is expected to lead to the largest noise contribution.

We choose the Model 2 and 4, and the x-variable of the Lorenz system as examples of this influence. Model 2 is linear, Model 4 is nonlinear, both with memory over 1 time step, and the Lorenz system has infinite memory, at least in theory. We use the same noise settings for each of these systems as described earlier.

\begin{table}
\caption{Causal strengths and certainties for different models as function of reference density}
\centering
\begin{ruledtabular}
\begin{tabular}{cccccccccc}
&  		& Model 2   & 		&  	& Model 4 	& 	& 	& Lorenz 		&  \\
&		&		&			&		&			&	&	&x-variable 	&  \\
\textbf{Estimate} & Cauchy &Gaussian& Uniform & Cauchy & Gaussian & Uniform & Cauchy & Gaussian & Uniform	 \\
\hline
$W(x)$ 				& 0.30	& 0.00 	& 0.66	& 0.28	& 0.24	& 1.39  	& 0.39	& 0.04	& 0.21 \\
$W(x|y,z)$ 			& 2.96 	& 2.66	& 3.30	& 2.76 	& 2.71	& 3.87 	& 4.35 	& 4.01	& 4.15\\
$cs(x^{n+1},y^n)$			& 0.06 	 & 0.06	& 0.05	& 0.48 	 & 0.49	& 0.34 	 & 0.27 	 & 0.30	& 0.29\\
$cs(x^{n+1},z^n)$  			& 0.84 	 & 0.94	& 0.76	& 0.41 	 & 0.42	& 0.30 	 & 0.15 	 & 0.16	& 0.16  \\
$cs(x^{n+1},x^n)$			&  		 & 		&  		&  		 & 		&   		& 0.49 	 & 0.53	& 0.51\\
$cs(x^{n+1},\eta^n)$			& 0.10 	& 0.00	& 0.20	& 0.10 	 & 0.09	& 0.36   	& 0.09 	 & 0.01	& 0.05 \\

\end{tabular}
\end{ruledtabular}
\end{table}

The results are shown in Table VII. The first and the last row, the latter a normalization of the first, show the same trend for models 2 and 4. The estimated noise contribution is lowest for the Gaussian reference density, the highest for the uniform density, and the results for the Lorentz-Cauchy density are in between. For model 2, in which the target is Gaussian distributed, the noise contribution using the Gaussian reference density is indeed zero. The Lorenz 1963 model behaves differently in that the largest estimated noise contribution comes from using the Lorentz-Cauchy density. This is not surprising as the strange attractor of that system has extremes which do not vary much from one realization to the other, so the uniform density is closer than the Lorentz-Cauchy density with its wide tails, see figure 7. 

\begin{figure}
\centering
\includegraphics[width=0.5\linewidth]{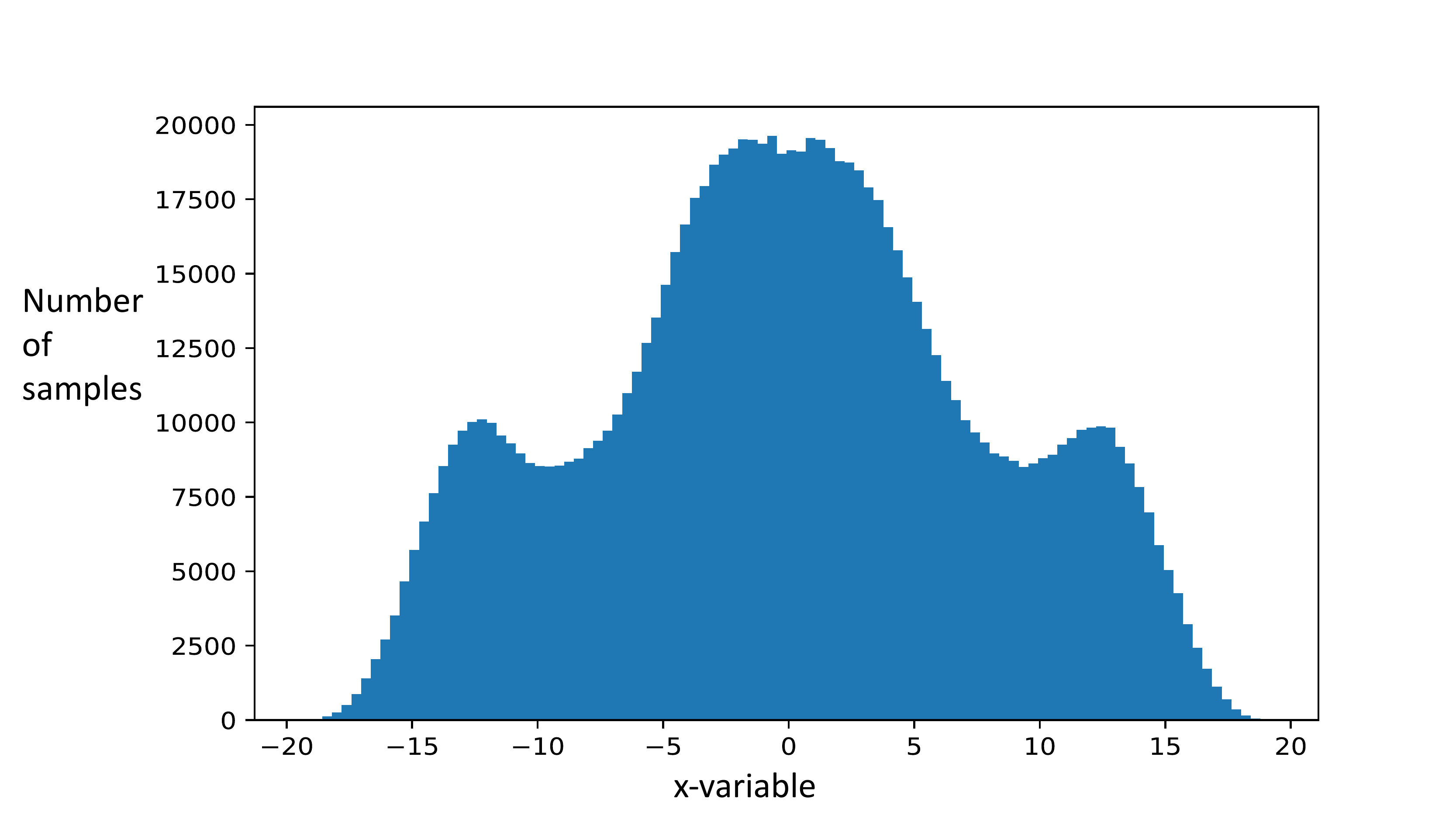}
\caption{Histogram of x-variable from 1,000,000 samples from a Lorentz 1963 model simulation. Note the finite size of the domain and image of this model.}
\end{figure}   

These numbers show that uniform density is most conservative estimate of how much the drivers explain about the target, and might be considered the preferred reference density for model 2 and 4. However, the boundaries of the uniform density are determined by the two extremes in the target time series, and hence can be strongly dependent on the actual realization of the noise. (As discussed, a counter example is the Lorenz 1963 system, which has a very strong attractor, but one would not expect such behavior in general.) Strong dependence on the noise realization is an undesirable property of the uniform density. Because the structure of the Lorentz-Cauchy density is determined by averaged quantities of the target time series, its mean and its entropy, it is less sensitive to a specific realization of the noise. Furthermore, as mentioned before, the Lorentz-Cauchy density has extremely wide tails, and is in that sense closest to a wide uniform density.

We stress again that if the present framework is used one has to specify the reference density. For discrete variables the natural reference density is the uniform density, and the discussion presented here does not apply. 
 
\subsection{Application to ENSO}
The last example is a real-world example based on time series of monthly indices representing the El-Nino-Southern Oscillation (ENSO) phenomenon from 1979 until 2019. The time series are  the sea-surface temperature in Eastern tropical Pacific (Nino3.4, called the NINA34 time series), the tropical east-west wind at a height of $850\;hPa$ (U850), the tropical east-west wind at $200\;hPa$ (U200), and the ocean heat content (HEAT, upper 300 m). The data have been extracted from https://psl.noaa.gov/enso/dashboard.lanina.html where details on their exact meaning and generation can be found.
 To reduce the influence of noise we perform a 5 point moving average on each time series.

 We want to infer the cause for the sea-surface temperature (SST) of the Eastern Pacific, which we measure with the Nino3.4 index. The complication is that several processes can influence Nino3.4 at different time lags, including Nino3.4 itself. Since the time series is relatively short, only 492 monthly time steps, we cannot explore the new framework in full. Each variable at a different lag would be a process in the new framework, so if we assume up to 6 months time lag we would have $4 \times 6 = 24$ processes. It is impossible to find accurate estimates of high-dimensional integrals with only 492 time points. On the other hand, geophysical systems often display different causal structures operating on different time scales. To solve this problem we concentrate on the influence of all 4 drivers on this target with a time lag between 3 and 4 months. This choice stems from a calculation of that time lag between the target and each of the drivers for which the mutual information between them is largest. It turns out that U200 has a largest mutual information with the target at 1 month lag, U850 at 2-3 months, and HEAT at 3-4 months. We choose the largest of these as that is most interesting in terms of long-term prediction.

Table VIII shows the total causal strengths for each driver and a decomposition in the 1links, 2links, etc.. We first note that all 4 processes are causal to Ni\~no3.4 at the 3-4 months time lag, and the most important one is HEAT, so the ocean heat content in the upper $300\;m$. This is understandable as the heat content is related to the down-welling Kelvin wave that sets off an El-Nino, or the upwelling Kelvin wave that sets off a La-Nina. Such a wave takes about 3 months to travel from the West to the East equatorial Pacific. The 1links are much smaller than the total causal strengths, pointing to the importance of the interactions between the variables to drive the target. It is through these interactions with others via 2links and 3links that the 1link of each driver is increased by a factor of order 2 to the total causal strength. This makes perfect sense physically because ENSO is a strongly coupled ocean atmosphere system with strong feedbacks.

 To analyze this further we show a full decomposition of the causal web in figure 8. The numbers in the circle denote the 1links, the blue lines denote the 2links, the red lines the 3links, and the 4link is depicted in green. The connection between these numbers and Table VIII is as follows. To find the total 2link contribution from e.g. HEAT in Table VIII we add all its 2 link values and divide the result by 2. As explained earlier, the division by two denotes that each 2link in figure 8 has to be divided over the two contributing processes. This leads to $(7+6+3)/2 = 8$. (Note that rounding errors appear when not all digits available are used.) 3links are obtained in a similar way with division by 3, etc. Note that the target itself is not displayed in this figure.

The largest 2 links are between HEAT and NINA34 (7\%) and HEAT and U850 (6\%). The first shows that a high SST together with a high heat content lead to a high SST 3-4 months later. The second shows that a strong positive wind anomaly pushes the heat anomaly further East, again enhancing the SST there. The 3link between these three processes shows that on top of these 2link interactions they also work in concert to influence the NINA34 SST 3-4 months later. This is because a higher SST will strengthen the U850 via wind convergence above the high SST,  which will enhance the SST 3-4 months later via an ocean heat content anomaly driven by this wind. Remember that a 3link is the interaction term between the 3 drivers with the 2links and the 1links subtracted. Hence, this 3link indeed denotes a physical connection between the 3 drivers.

Of note is the low interaction of U200 with the other processes: it does interact via 2links with NINA34 and HEAT at a 2-3\% level, and the 2link with U850 is only 1\%. Its 3link with Heat and U850 is even negative, showing that these 3 processes together reduce the causal relation between each of them and the target. However, the 4link between all 4 drivers is relatively large at 11\%. This means that if we add NINA34 to this trio the causal relation with the target is enhanced. This is understandable because the connection between what happens in the ocean-atmosphere boundary layers is connected to the upper atmosphere via the SST. The SST drives of suppresses vertical advection and hence the connection to U200. It is only in interaction with HEAT and low-level winds that this process can influence the evolution of El NINO, and the SST 3-4 months in advance.

Finally, we checked for missing processes by perturbing the target time series by random Gaussian noise with standard deviation 0.05, to be compared with the standard deviation of 0.7 for the Nina34 index itself. The actual error in the index is unknown, so we took a noise standard deviation value that was almost 10\% of that of the signal. This resulted in a causal contribution of the noise term in the framework of 0.220, against 0.219 for the unperturbed target, a change of less than 1\%. This insensitivity to observational noise shows that there are important missing processes in the system, as explained in section V. This is not surprising as we know that the ENSO phenomena is a complicated coupled climate mode that can only partly be described by the 4 drivers we used here.

In comparison with other frameworks, we note that Transfer Entropy will not provide information on the 2-,3-, and 4links, missing out significantly on the physics. We have not found a useful comparison with PID, as its component terms are hard to define. However, even if we managed to do that PID does not decompose the Synergistic and Redundant terms, missing out on the physical interpretation.

CCM determines binary causal relations, and hence it is also not able to disentangle the richness of the underlying physics. We used a band-pass Butterworth filter of order 4 and band periods 0.05-0.35 months. The optimal lag and embedding dimension were found by trial and error for HEAT, U850 and  U200 with Nina34. Figure 9 displays the results (note that the arrows in the legend indicate driver to target, not predictive power). The strongest driver for Nina34 is HEAT, followed by U200. U850 doesn't seem to converge, suggesting no (strong) driving from U850. The CCM score indicating Nina34 as driver suggests that the causal relations tend to be bidirectional. However, the converge of all lines is rather weak, suggesting that a longer time series is needed. Furthermore, noise seems to hamper causal identification. It is hard to infer physical relations from these curves.

\begin{table}
\caption{Mlink values and total strength in \% for target Nino3.4}
\centering
\begin{ruledtabular}
\begin{tabular}{cccccc}
\textbf{Estimate}& 1links & 2links	& 3links & 4links & causal strength \\
\hline
HEAT & 	11 & 8 & 3 & 3 & 25 \\
Nina34 & 9 & 6 & 4 & 3 & 22 \\
U850 & 8 & 5 & 3 & 3 & 19 \\
U200 & 6 & 3 & 1 & 3 & 13 \\
'noise'	& & & &		& 22 \\
\end{tabular}
\end{ruledtabular}
\end{table}

\begin{figure}
\centering
\includegraphics[width=0.7\linewidth]{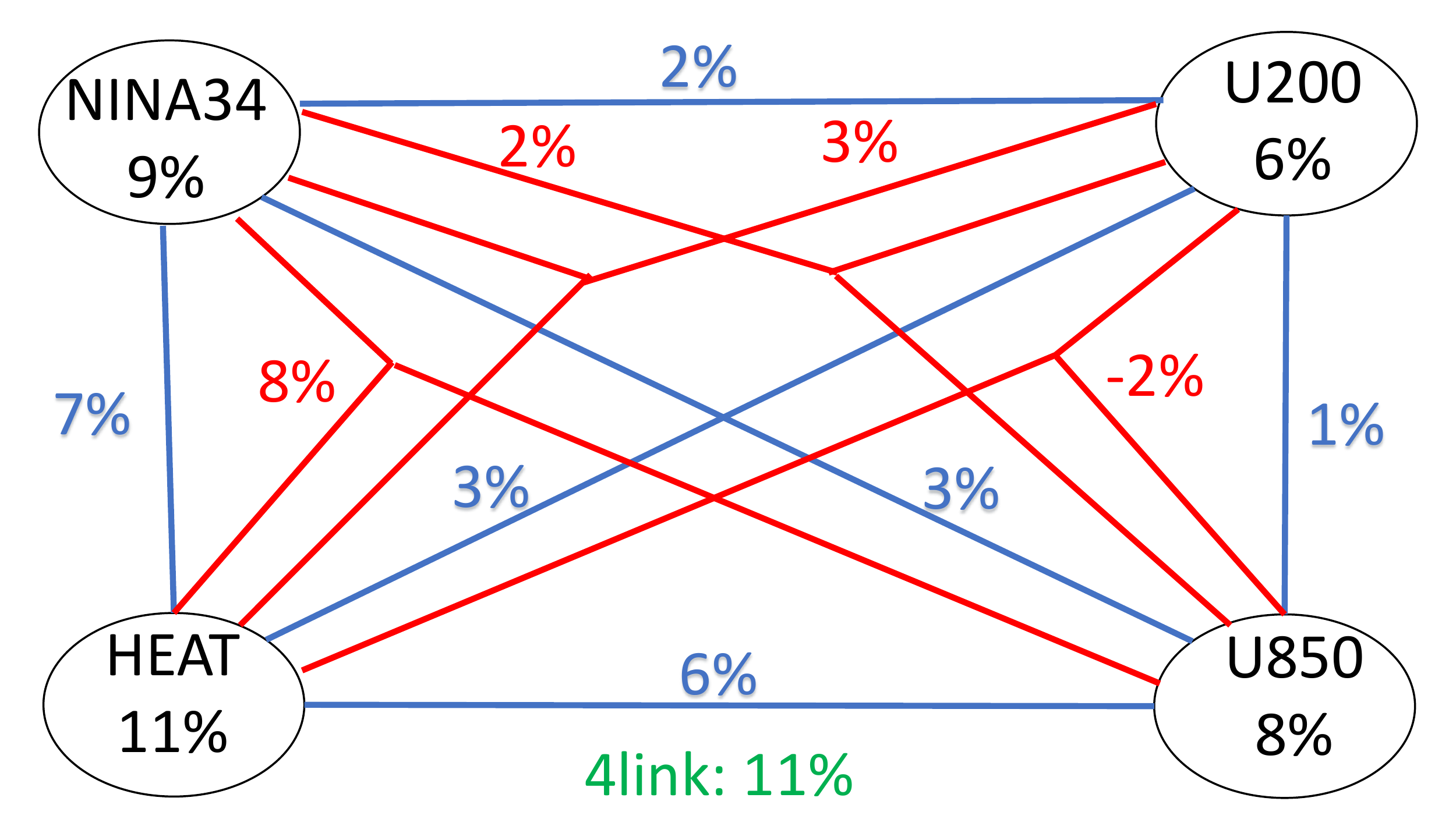}
\caption{Causal net for drivers of Nino3.4 using lags 1 to 4 months for each driver (see text). The circles denote the drivers and their 1link contribution, the blue lines are the 2links, the red lines the 3links, and the 4link is the same for all in green.}
\end{figure}  

\begin{figure}
\centering
\includegraphics[width=0.7\linewidth]{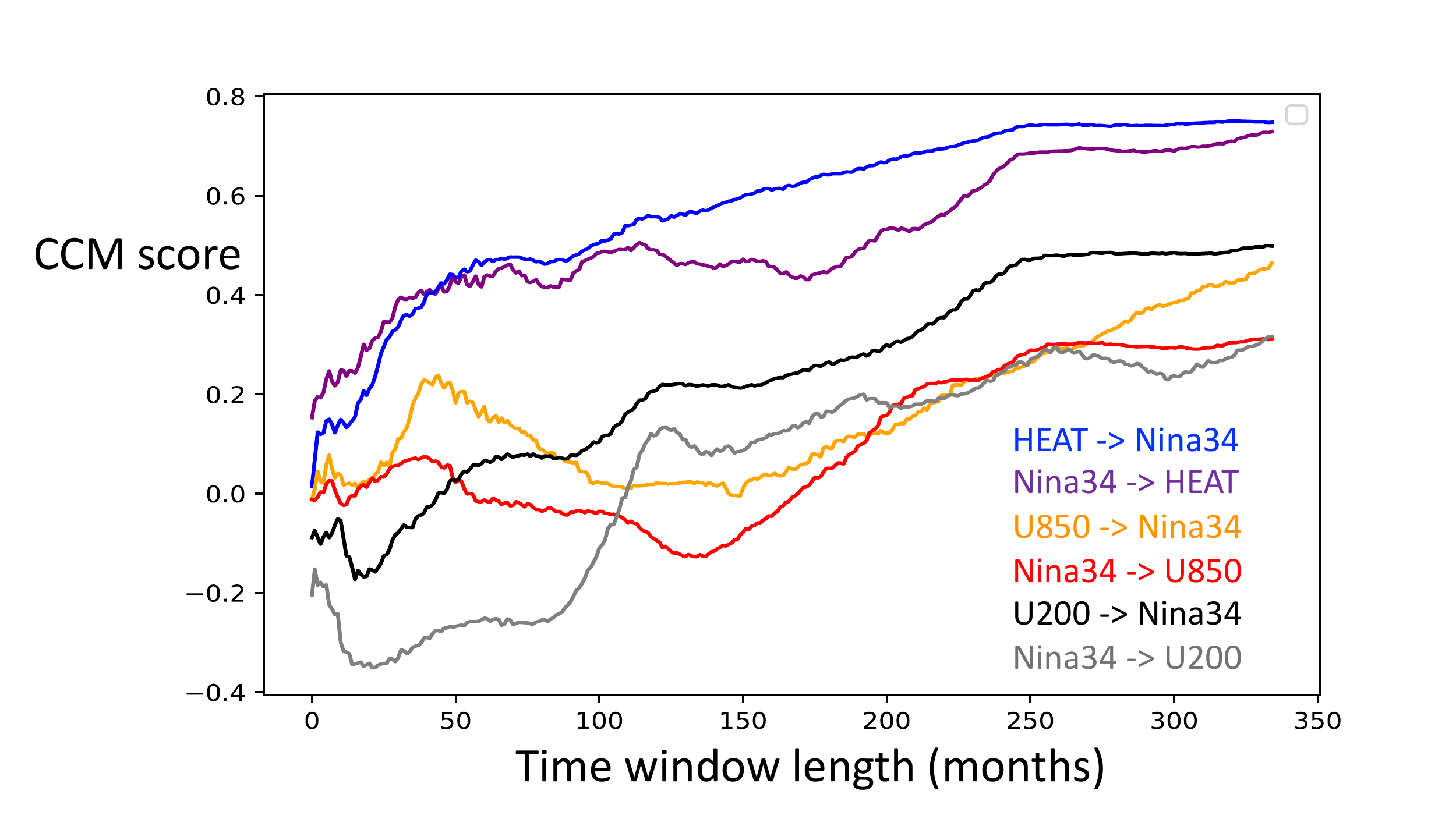}
\caption{CCM-scores indicating driver -> target relations, versus the length of the time series used. Note that the orange and grey lines do not seem to converge, suggesting no causal relation.}
\end{figure}  
\section{Discussion}

A new causal discovery framework has been developed based on a complete decomposition of total mutual information between a target process and all its potential drivers. It builds on certainty, a non-negative quantity and invariant under nonlinear uni-variable transformation, unlike differential entropy. This allows us to infer how knowledge of driver processes increases our
knowledge of a target process, so how it increases our certainty about that process. 
We can decompose the contribution of each driver process in direct contributions, 
and joint contributions between 2 processes, between 3 processes etc. 
This decomposition is rich as it allows a detailed characterization of the underlying causal structure.
By normalizing each contribution different studies can be compared, and the self-certainty
can be reinterpreted as the contribution from unknown processes, allowing us a quantification of the
processes not included in the causal discovery set, including the importance of confounders. In this sense
the framework can be considered a complete framework.

We showed in simple dynamical systems the advantage of including the joint contributions 
over traditional approaches. Using the Lorenz 1963 system as an example, we showed that
the framework will, via the causal strengths, contain information about the governing equations, while 
the 1links and 2links reveal information
on the underlying low-dimensional structure that the dynamics live on. 
In the Lorenz 1963 example these links reveal features of the strange attractor, and even the dynamics on that strange
attractor. Furthermore, using the framework on real-world timeseries of ENSO indices we showed a tight coupling between the resulting causal web and the underlying physics of the system.

The framework has a few drawbacks that need discussing. For continuous variables
we need to define a reference probability density as function of the target variable. The resulting causal strengths do depend on this density.
A thorough investigation of several possibilities lead us to conclude that the Lorentz of Cauchy density has many advantages compared to others, 
and is the density of choice in this paper. Since any result obtained with one reference density can be transformed to those using another 
reference density, the main message is that the reference density used should be reported with the causal strength values.

Another potential drawback is the number of calculations involved.
In general, when there are $N$ driver processes, the number of (conditional) mutual informations that need to be calculated is $2^N$.
Often, however, a large number of the driver processes is related to connections at larger time lags. Assumptions on the structure of the underlying system, e.g. 1st-order Markov,
would make many of these mutual informations non causal, reducing the number of calculations needed.
As an example from the
Lorenz 1963 system, the direct 1link contributions more than 1 time step back are all zero because the conditioning
blocks the information: $I(x^{n+1};y^{n-1} | x^{n},y^{n},z^{n},...) = 0$. Similar remarks hold for higher-order links
and can be generalized as follows for a 1st-order Markov system: All conditional mutual informations that condition on all variables at the same time will block
information flow from before to after that time. Extensions like this can be made for 2nd-order Markov processes, etc.
The point is that if more is known about the underlying dynamics we can use that to reduce the number of calculations needed.
As a final remark on calculations, since all (conditional) mutual information calculations are independent of each other the causal 
calculations are highly efficient on parallel computer platforms. 

The framework is based on information theoretic measures such as mutual information. As has been known for some time, and e.g. \cite{James2017} showed convincingly, there are systems that have different internal dependencies but for which all information-theory based measures are identical. This means that we will not be able to see those internal dependencies with our framework. This, of course, is not surprising as entropy-based measures are integrals over nonlinear functions of the underlying probability density functions, and hence details of these probability density function will be lost. In fact, the argument can easily be pushed further to something like: any causal theory that relies on integral quantities of probability density functions will miss out on certain details in these densities, and hence potentially miss important causal structures. In our view it is impossible to avoid this issue as any causal theory is ultimately based on summary statistics. It is unknown what real-world causal structures are, but we do know that many systems do differ in entropy-based measures, and it is these systems that we intend to study with the present framework. 

An important ingredient of this framework is still missing: a proper uncertainty estimate on all terms. 
If long time series are available, one can split these up into shorter time series and calculate the sample variance
in the resulting sample of mutual information calculations. A handle on the bias could be obtained by 
using sub series of different length and compare sample means of different time series length calculations.
We are working on a complete Bayesian setting for the framework to accommodate this shortcoming
as hypothesis testing on zero causal strength, which is often used in present-day causal studies, is clearly not enough for scientific exploration.

For some realistic systems, such as financial time series or climate change time series, causal discovery needs to be assessed on  non-stationary time series. The presented framework would need to be extended to included them. There are several challenges to address. These include defining the main changes, which may be a function of the time scale of interest and the the length of the time series. As an example from climate science, we know climate, e.g. defined as the joint pdf of system Earth over a 30 year time scale, is changing. 
Performing time-series-based causal discovery over a 100 year time scale has to proceed with care for this pdf is changing. However, over a 
million-year time scale meaningful causal discovery can be performed treating the timeseries as stationary. As another example, systems with regime shifts 
can be treated as non-stationary, unless one wants to study the cause of the regime shifts. A promising venue for capturing causality on non-stationary time series  based on time-lagged information measures  has been proposed by \cite{Panapa2016}. They use rank vectors based on delay vectors from the time series to estimate the partial symbolic transfer entropy. 

Finally, although the present-day formulations such as PID and Convergent Cross Mapping have shortcomings it is important to better understand
what synergy and redundancy and unique contributions actually mean, and come up with a  closed 
system such as the framework presented in this paper, incorporating those ideas.


%
%

%

\begin{acknowledgments}
This work was funded through the European Research Council project CUNDA number 694509.
\end{acknowledgments}

\section*{Data Availability}
The data that support the findings of this study are available from the corresponding author
upon reasonable request.

\bibliography{causalbib.bib}

\end{document}